\documentclass[twocolumn]{IEEEtran}
\usepackage{amssymb}
\usepackage{amsmath}
\usepackage[version=4]{mhchem}
\usepackage{graphicx}
\usepackage{multirow}
\usepackage{wrapfig}
\usepackage{color}
\usepackage{relsize}
\usepackage{colortbl}
\usepackage{bm}
\usepackage{soul}
\usepackage{subfigure}
\usepackage{enumerate}
\usepackage{paralist}
\usepackage{dsfont}
\usepackage[normalem]{ulem}
\usepackage{mathtools}
\usepackage{tabularx,booktabs,caption}
\usepackage[flushleft]{threeparttable}\usepackage{tabularx,booktabs,caption}
\usepackage[flushleft]{threeparttable}

\newcommand{\p}{\mathrm{p}}

\newcommand{\E}{\mathrm{E}}
\newcommand{\e}{\mathrm{e}}
\newcommand{\Var}{\mathrm{Var}}

\newcommand{\Cov}{\mathrm{Cov}}
\newcommand{\MSE}{\mathrm{MSE}}
\newcommand{\NMSE}{\mathrm{NMSE}}
\def\mathclap#1{\text{\hbox to 0pt{\hss$\mathsurround=0pt#1$\hss}}}
\makeatletter
\newcommand{\vastt}{\bBigg@{3}}
\newcommand{\vast}{\bBigg@{4}}
\newcommand{\Vast}{\bBigg@{5}}
\makeatother

\DeclareMathOperator*{\argmin}{arg\,min}

\begin{document}
\title{Channel Sensing in Molecular Communications with Single Type of Ligand Receptors }
\author{Murat Kuscu,~\IEEEmembership{Student Member,~IEEE}
        and Ozgur B. Akan,~\IEEEmembership{Fellow,~IEEE}
       \thanks{The authors are with the Internet of Everything (IoE) Group, Electrical Engineering Division, Department of Engineering, University of Cambridge, Cambridge, CB3 0FA, UK (e-mail: \{mk959, oba21\}@cam.ac.uk).}
       \thanks{Ozgur B. Akan is also with the Next-generation and Wireless Communications Laboratory (NWCL), Department of Electrical and Electronics Engineering, Koc University, Istanbul, 34450, Turkey  (email: akan@ku.edu.tr).}
	   \thanks{This work was supported in part by the ERC projects MINERVA (ERC-2013-CoG \#616922), and MINERGRACE (ERC-2017-PoC \#780645).}}


\maketitle

\begin{abstract}
	Molecular Communications (MC) uses molecules as information carriers between nanomachines. MC channel in practice can be crowded with different types of molecules, i.e., ligands, which can have similar binding properties causing severe cross-talk on ligand receptors. Simultaneous sensing of multiple ligand types provides opportunities for eliminating interference of external molecular sources and multi-user interference (MUI), and developing new multiple access techniques for MC nanonetworks. In this paper, we investigate channel sensing methods that use only a single type of receptors and exploit the amount of time receptors stay bound and unbound during ligand-receptor binding reaction to concurrently estimate the concentration of multiple types of ligands. We derive the Cram\'er-Rao Lower Bound (CRLB) for multi-ligand estimation, and propose practical and low-complexity suboptimal estimators for channel sensing. We analyze the performance of the proposed methods in terms of normalized mean squared error (NMSE), and show that they can efficiently estimate the concentration of ligands up to $10$ different types with an average NMSE far below $10^{-2}$. Lastly, we propose a synthetic receptor design based on modified kinetic proofreading (KPR) scheme to sample the unbound and bound time durations, and a Chemical Reaction Network (CRN) to perform the required computations in synthetic cells.
\end{abstract}

\begin{IEEEkeywords}
Molecular communication, receiver, ligand receptors, channel sensing, multi-molecule sensing, maximum-likelihood estimation, method of moments, multiplexed detection, molecular division multiplexing, molecular division multiple access, kinetic proofreading, synthetic biology, chemical reaction networks.
\end{IEEEkeywords}

\IEEEpeerreviewmaketitle

\section{Introduction}
\label{sec:Introduction}
\IEEEPARstart{M}{olecular} Communications (MC) is a bio-inspired communication technique based on using molecules to encode, transmit and receive information \cite{akan2017fundamentals}. MC has gained increasing popularity in the last decade due to its potential to enable artificial nanonetworks, i.e., networks of nanomachines, and the Internet of Nano Things (IoNT), an emerging technology consisting in nanonetworks connected to the Internet and promising for unprecedented applications, e.g., intrabody continuous health monitoring for early diagnosis and treatment, smart drug delivery, and artificial organs \cite{akyildiz2010internet, akyildiz2015internet, kuscu2016internet}. MC has been extensively studied from various theoretical aspects, e.g., detection, channel modeling, and modulation \cite{kuscu2018survey}; however, there is still no practical implementation of an MC system at micro/nanoscale, except for a few microscale testbeds developed for bacterial MC \cite{nakano2008microplatform, krishnaswamy2013time, grebenstein2019biological}. The key to realize nanoscale artificial MC networks is to close the gap between the theory and practice by addressing the peculiarities resulting from discrete nature of molecules, limited capabilities of nanoscale devices, and highly stochastic, nonlinear, and time-varying dynamics of the MC channels, which bring about new challenges fundamentally different from those we tackle in conventional electromagnetic (EM) wireless communications. 
 
In practice, MC channels, especially in physiologically-relevant environments, can be crowded by many types of molecules that may have similar characteristics rendering their discrimination nontrivial. These molecules can be resulting from natural processes, e.g., intrabody cell signaling, that are generally not relevant to the MC application, leading to natural interference. They can also result from another MC system co-existing in the same medium leading to multi-user interference (MUI) \cite{dinc2017theoretical}. The knowledge of the channel state information (CSI) in terms of instantaneous concentration of co-existing molecules is crucial for developing reliable detection and modulation methods in the time-varying presence of interferer molecules of similar characteristics with the messenger molecules. It can also enable the application of cognitive medium access (MA) techniques for MC nanonetworks to efficiently allocate the limited molecular resources \cite{alizadeh2018coexistence}. Channel estimation techniques are proposed for MC with passive and transparent observers in \cite{jamali2016channel, noel2015joint}, and channel sensing techniques for co-existing MC nanonetworks that utilize the same type of molecules are considered in \cite{alizadeh2018coexistence}. However, simultaneous estimation of concentration of different molecule types has not been investigated in the MC literature before.  

This study is focused on synthetic biological MC transceivers with ligand receptors on their surface. The receptors constitute the interface between the exterior and interior of the living cells, interact with the external ligands (i.e., molecules in the MC channel, based on ligand-receptor binding reaction), and transduce the binding events into intracellular signals \cite{bialek2012biophysics}. The binding rate of ligands to the receptors depend on the transport properties and the concentration of ligands as well as the activation energy of the ligand-receptor binding reaction, whereas the unbinding rate only reflects the affinity between the ligands and the receptors at equilibrium \cite{pan2013molecular}. Ligand receptors, in practice, can provide specificity for the target ligands only to a finite extent, as other types of ligands can also bind the same receptors, though typically having lower affinities \cite{mora2015physical}. The correlation between the unbinding rate and the ligand type is the key property that is exploited in this paper to develop molecular channel sensing methods. 

MC with ligand receptors has been addressed from different aspects. Channel models are developed between point transmitters and reactive receivers with ligand receptors in \cite{deng2015modeling, ahmadzadeh2016comprehensive}. Detection techniques are proposed for concentration shift keying (CSK) modulated MC with ligand receptors in \cite{kuscu2016modeling}, based on sampling the instantaneous number of bound receptors. Recently, the continuous history of bound and unbound states has proven to provide more information about the external ligand concentration \cite{chou2015markovian, endres2009maximum}. In this direction, maximum a posteriori (MAP) detection methods are proposed for MC in \cite{chou2015markovian, chou2015maximum}. In our previous work \cite{kuscu2018maximum}, receptor unbound time duration statistics is shown to provide a larger dynamic input range for the detector to cope with saturation at high ligand concentrations resulting from intersymbol interference (ISI). Maximum likelihood (ML) estimation of the concentration of two different ligand types based on sampling the receptor bound time durations is studied in \cite{mora2015physical, singh2017simple, lalanne2015chemodetection, siggia2013decisions}. These studies also argue practical implementations of the ML estimators exploiting kinetic proofreading (KPR) mechanism, which is an active cellular mechanism that increases specificity, suggested to exist in T-cell receptors that can sense very low concentrations of foreign agents with extreme specificity as part of the immune system \cite{mckeithan1995kinetic}. Based on a similar approach, we have previously investigated the performance of ML detection with bound time durations in the presence of single type of interferers, and shown that it outperforms other MC detection schemes that use the samples of receptor unbound time or instantaneous receptor states \cite{muzio2018selective}. However, none of the previous studies has considered the problem of sensing the concentration of more than two types of ligands at the same time. 

In this paper, we develop practical channel sensing methods to concurrently estimate the concentrations of different ligand types co-existing in the channel. Our work draws on the recent biophysics literature \cite{mora2015physical, singh2017simple}, that exploit the receptor cross-talk to sense two types of ligands with a single type of receptors, and generalizes it to the concentration estimation of more than two ligand types for MC applications. The proposed channel sensing methods consist of two estimators. The first one is an unbiased maximum-likelihood (ML) estimator of the total ligand concentration, which uses the amount of time the receptors stay unbound, as proposed in \cite{endres2009maximum, kuscu2018maximum}. This estimator exploits the fact that as the total ligand concentration increases, the receptors bind the ligands more frequently, and thus, the unbound periods of the receptors get shorter. The second estimator is also unbiased, and estimates the concentration ratio of each ligand type from the amount of time the receptors stay bound to a ligand, using the method of moments (MoM). This estimator is based on the fact that the receptors bind more strongly to the ligands of higher affinity, and thus, the duration of bound time periods are correlated with the type of ligands \cite{mora2015physical}. We also develop a more practical version of the concentration ratio estimator, which is biased, however, requires less number of computations. The product of the total concentration and the ratio estimators provides the instantaneous concentration of each ligand type. We evaluate the performance of the channel sensing methods in terms of normalized mean squared error (NMSE) averaged over all co-existing ligands, for varying number of ligand types in the mixture, varying number of samples, and similarity between the ligand types, and varying concentration distributions of different ligand types within the mixture.  

The estimators should operate inside synthetic cells by making use of second messengers, i.e., intracellular signaling molecules, for arithmetic calculations. This requires the transduction of unbound and bound time durations into concentration of second messengers, which are then processed through analog computing. To this end, we propose a synthetic receptor design with a multitude of internal states, that utilizes a modified version of the conventional KPR mechanism \cite{mckeithan1995kinetic}. The proposed receptor is able to be activated by an intracellular activation signal at the start of a sampling period, and encode the observed unbound and bound time durations into the concentration of different types of second messengers. Lastly, we discuss the implementation of the channel sensing methods in synthetic cells, and propose a Chemical Reaction Network (CRN)-based approach to realize the required computations.

The remainder of this paper is organized as follows. In Section \ref{sec:motivation}, we discuss the opportunities of multi-molecular channel sensing in MC focusing on its potential in developing reliable detection methods, adaptive and multi-functional receivers, and medium access techniques. In Section \ref{sec:statistics}, we review the fundamental properties of ligand-receptor binding reactions. We present the mathematical framework of the proposed channel sensing methods in Section \ref{sec:spectrumsensing}. The performance of the technique is evaluated in Section \ref{sec:performance}. We provide a practical discussion on implementation of the proposed method in Section \ref{sec:discussion}. Lastly, we conclude the paper in Section \ref{sec:conclusion} by discussing open research directions. 

\section{Opportunities of Channel Sensing in MC}
\label{sec:motivation}
Exploiting the cross-talk between different types of ligands for multi-molecular channel sensing with single type of receptors is crucial for improving the adaptivity and reliability of the MC devices, increasing the capacity of the MC channels, and enabling the effective use of limited molecular resources for medium access without requiring substantial amount of additional computational resources and receptors. The proposed channel sensing methods can prove effective especially towards the following directions in the MC research: 
\begin{itemize}
\item \textbf{Development of reliable detection methods for CSK modulated signals based on eliminating the interference of similar ligands released by external sources:} Current studies focusing on CSK-based MC with ligand receptors assume that the receptors are ideal, such that they only react with the ligand type that carries the information \cite{kuscu2018maximum}. However, in practice, the specificity of receptors is not perfect, and they can react with multiple types of molecules, though with different reaction rates, especially in physiologically relevant conditions. Eliminating the interference by sampling the instantaneous receptor states is not viable, when the channel is time-varying, e.g., the concentration of interferer molecules change between signaling intervals. The channel sensing methods proposed in this paper do not require a priori knowledge of the probability distribution of ligand concentrations; therefore, it can enable robust and reliable detection under time-varying conditions. 

\item \textbf{Development of reliable detection methods for molecule shift keying (MoSK) and ratio shift keying (RSK) modulated MC signals:} These modulation techniques rely on the transmission of multiple types of ligands. In MoSK, the information is encoded into the concentration of different ligand types, which are transmitted in separate signaling intervals \cite{kuran2011modulation}. On the other hand, in RSK, the information is encoded into the ratio of concentrations of different ligand types transmitted at the same signaling interval. The current studies focusing on both modulation methods assume that there is an ideal receptor for each ligand type, and the cross-talk between different ligand-receptor pairs is neglected \cite{shahmohammadian2012optimum, kim2013novel}. However, this is not the case in practice, as the cross-talk between ligands always exists. The proposed channel sensing methods can be employed to eliminate the cross-talk between different ligand-receptor pairs and increase the capacity of the channel. Additionally, with the use of the proposed method, both MoSK and RSK modulated MC signals can be accurately detected by utilizing only a single type of receptor. This can also enable the transmitter to increase the cardinality of the set of transmitted molecule types for boosting the channel capacity, without necessitating the deployment of extra receptors in the receiving side. 

\item \textbf{Development of interference-free molecular division multiple access (MDMA) techniques:}  MDMA is based on the idea of using different types of molecules in different MC channels co-existing in the same environment \cite{gine2009molecular, kuscu2016physical}. In this way, multiple nodes can concurrently use the same medium for information transmission; however, the MUI cannot be completely avoided, as the specificity of the receptors is not infinite. Moreover, the number of different  ligand types that can be generated and detected by resource-constrained MC devices is limited. Biocompatibility concerns for \textit{in vivo} applications make this limitation more severe  \cite{alizadeh2018coexistence} In these circumstances, as similar to the cognitive radio techniques studied for conventional EM communications \cite{akan2009cognitive}, the channel sensing methods can be opportunistically used to dynamically sense the utilization of different types of carrier molecules in the channel, prior to transmission, to avoid crowding the medium with a particular type of molecule and degrading the communication performance. On the receiver side, the multi-molecule channel sensing methods can provide the receiver with the required adaptivity in detecting different types of molecules transmitted. This also enables the receiving node to simultaneously communicate multiple transmitting nodes through molecular division multiplexing by preventing cross-talk from affecting the reliability of the communication.  

\item \textbf{Multi-functionality:} Lastly, the proposed technique can enable multi-functional MC devices that can simultaneously perform communication and sensing of multiple types of molecules using the same receptors. This can also help reduce the energy and molecular costs, and simplify the design of biological MC devices for MC nanosensor network applications. 
\end{itemize}

\section{Statistics of Ligand-Receptor Binding Reactions}
\label{sec:statistics}

In ligand-receptor binding reaction taking place on the surface of a biological MC device, e.g., engineered bacteria, receptors randomly bind ligands in their vicinity. A receptor can be either in the Bound (B) or Unbound (U) state, with exponential dwell times depending on the binding and unbinding rates of the ligand-receptor pair. The state of a single receptor exposed to a concentration of \textbf{\emph{single type of ligands}} is governed by the following two-state stochastic process,
\begin{equation}
\ce{U  <=>[{c_L(t) k^+}][{k^-}] B},
\label{eq:bindingreaction}
\end{equation}
where $c_L(t)$ denotes the time-varying ligand concentration in the vicinity of receptors, $k^+$ and $k^-$ are the binding and unbinding rates of the ligand-receptor pair, respectively \cite{berezhkovskii2013effect}. Note that the transition rate from unbound to bound state is modulated by ligand concentration $c_L(t)$. In diffusion-based MC, due to the low-pass characteristics of the diffusion channel, the bandwidth of the $c_L(t)$ is typically significantly lower than the characteristic frequency of the binding reaction, i.e., $f_B = c_L(t) k^+ + k^-$; thus, the ligand-receptor reaction is usually assumed to be at equilibrium with a stationary ligand concentration, which we denote simply by $c_L$. We note that, with this assumption, the process in \eqref{eq:bindingreaction} becomes a Continuous Time Markov Process (CTMP). In equilibrium conditions, the probability of observing a receptor in the bound state is given as
\begin{equation}
p_B  = \frac{c_L}{c_L+K_D},
\end{equation}
where $K_D = k^-/k^+$  is the dissociation constant, which gives a measure of the \textit{affinity} between a ligand-receptor pair. If there are multiple receptors that are independently exposed to the same ligand concentration and not interacting with each other, the number of bound receptors becomes a binomial random variable with a success probability of $p_B$. Hence, the mean and the variance of the number of bound receptors can be written as 
\begin{align}
\E[n_B] &= \frac{c_L}{c_L + K_D} N_R, \\ \nonumber
\Var[n_B] &= p_B (1-p_B) N_R,  \label{eq:Binomial}
\end{align}
where $N_R$ is the total number of receptors. 

\begin{figure}[!t]
	\centering
	\includegraphics[width=9cm]{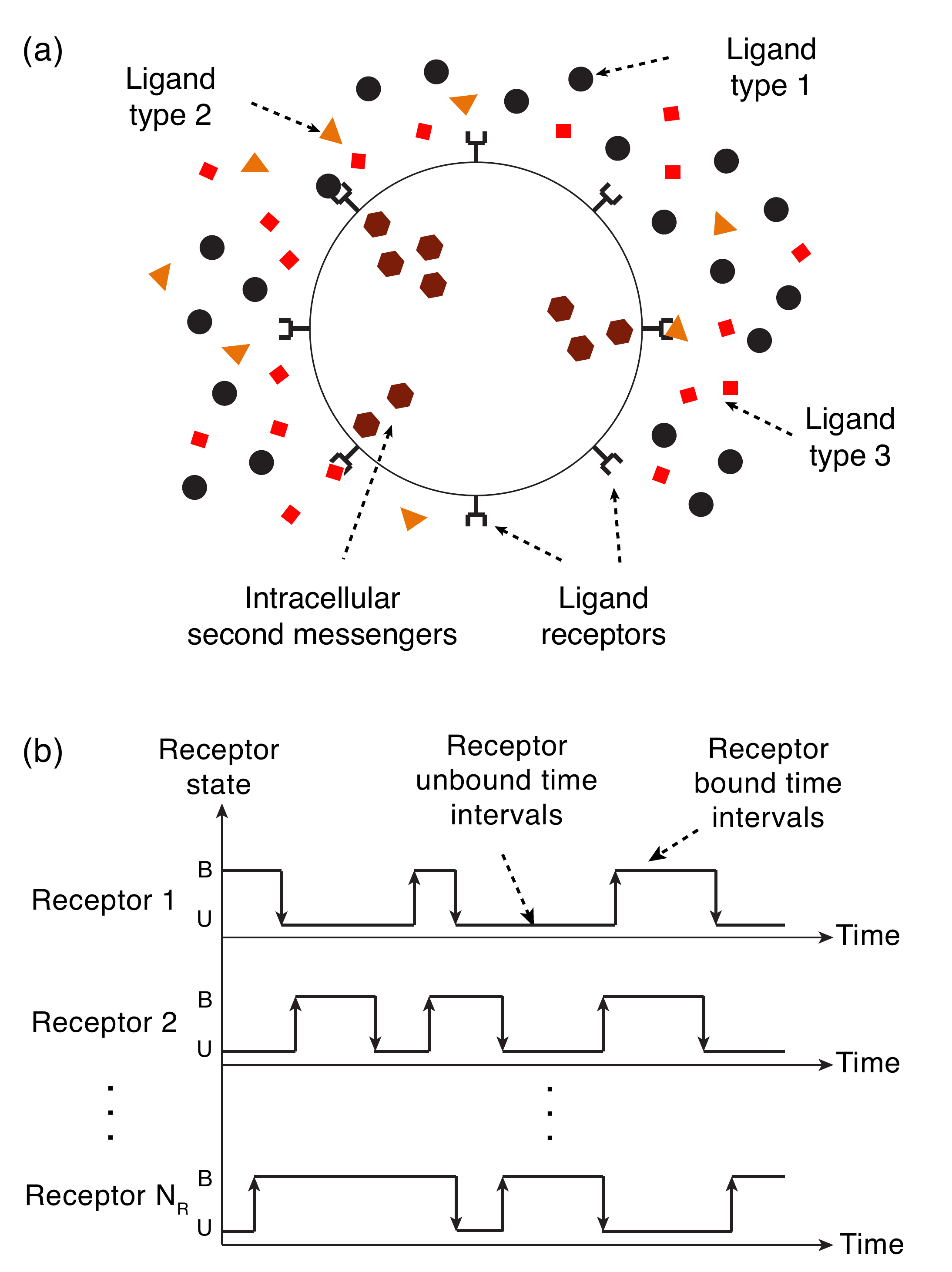}
	\caption{(a) A biological MC device with ligand receptors on its surface exposed to a mixture of different ligand types of different affinities with the receptors. The binding events are transduced into second messengers inside the cell. (b) An example time course of binding and unbinding events occurring on the receptors. The binding time durations change depending on the affinity of the bound ligand with the receptor. }
	\label{fig:receiver}
\end{figure}

Sampling the number of bound receptors at a given time instant previously proved effective in inferring the concentration of ligands, when the receiver is away from saturation \cite{kuscu2018maximum}, i.e., when $p_B \ll 1$. However, when there are \textbf{\emph{multiple types of ligands}} in the channel medium, as shown in Fig. \ref{fig:receiver}(a), which can bind the same receptor with different affinities, i.e., with different dissociation constants, the bound state probability of a receptor at equilibrium becomes 
\begin{equation} \label{prob_M}
p_B  = \frac{\sum_{i=1}^{M} c_i/K_{D,i}}{1 + \sum_{i=1}^{M} c_i/K_{D,i}},
\end{equation}
where $M$ is the number of different types of ligands co-existing in the medium. 

The expression in \eqref{prob_M} cannot be used to infer the individual ligand concentrations $c_i$ due to the interchangeability of the summands \cite{mora2015physical}. Therefore, in the case of a mixture of different ligand types, the required insight into the individual ligand concentrations can only be acquired by examining the continuous history of binding and unbinding events over receptors, which is exemplified in Fig. \ref{fig:receiver}(b). In this case, the likelihood of observing a set of $N$ independent binding and unbinding intervals over any set of receptors at equilibrium can be written as
\begin{align}\label{eq:likelihood1}
&\p\left(\{\tau_b,\tau_u \}_{N}\right) \\ \nonumber
&= \frac{1}{Z} \e^{-\mathlarger{\sum_{i=1}^N} \tau_{u,i} \left(\sum_{j=1}^{M} k_j^+ c_j \right)}  \mathlarger{\prod_{i=1}^{N}}  \left(\sum_{j=1}^{M} k_j^+ c_j k_j^- \e^{-k_j^- \tau_{b,i}} \right),
\end{align}
where $Z$ is the probability normalization factor, $M$ is the number of ligand types co-existing in the channel, $k_i^+$ and $k_i^-$ are the binding and unbinding rates for the $i^\text{th}$ ligand type, respectively, $\tau_{u,i}$ and $\tau_{b,i}$ are the $i^\text{th}$ observed unbound and bound time durations, respectively, \cite{mora2015physical, kuscu2018maximum}. We note that an unbound or bound time duration is the duration of a time interval that a receptor continuously stays unbound or bound, respectively. Given that the receptors are independent of each other, and they are exposed to the same ligand concentration assumed to be constant during sampling, ligand-receptor binding reaction becomes a stationary ergodic process. Therefore, the likelihood function \eqref{eq:likelihood1} does not depend on the time instants the ligands bind or unbind, and the indices of bound and unbound time durations do not necessarily imply a receptor-based or chronological order. In other words, the entire set of bound and unbound duration samples $\{\tau_b,\tau_u \}_{N}$ can be obtained equivalently by observing the time trajectory of only a single receptor or multiple independent receptors. 

In the diffusion-limited case, i.e., where the reaction rates are much higher than the characteristic rate of diffusion, the binding rate can be simply given by $k^+ = 4 D a$ for circular receptors \cite{mora2015physical}, with $D$ and $a$ being the diffusion constant of molecules and the effective receptor size, respectively. Assuming that the ligands are of similar size, their diffusion coefficients $D$, which depends on the temperature and viscosity of the fluid medium, and the size of diffusing molecules \cite{bialek2012biophysics}, are approximately equal for all ligand types. In this case, the likelihood function \eqref{eq:likelihood1} can be reduced to
\begin{align}
\p\left(\{\tau_b,\tau_u \}_{N}\right) = \frac{1}{Z} e^{- T_u  k^+ c_{tot}}  (k^+ c_{tot})^N \prod_{i=1}^{N} \p\left(\tau_{b,i}\right) ,
\label{eq:likelihood}
\end{align}
where $T_u = \sum_{i = 1}^N \tau_{u,i}$ is the total unbound time of all receptors, $c_{tot} = \sum_{i = 1}^M c_i$ is the total ligand concentration in the vicinity of the receptors, and $\p(\tau_{b,i})$ is the probability of observing a bound time duration, which is given as a \textit{mixture of exponential distributions}, i.e.,  
\begin{align}
\p\left(\tau_{b,i}\right)  =  \sum_{j=1}^M \alpha_j k_j^- \e^{-k_j^- \tau_{b,i}}.
\end{align}
Here $\alpha_j = c_j/c_{tot}$ is the concentration ratio of the $j^\text{th}$ ligand type. 

The log-likelihood function for an observed set of unbound/bound time durations can be written as the sum of three terms, i.e., 
\begin{align}\label{log_like}
\mathcal{L}(\{\tau_b,\tau_u \}_N) & = \ln\p(\{\tau_b,\tau_u \}_N), \\ \nonumber
& = \mathcal{L}_0  + \mathcal{L}\left(T_u | c_{tot}\right)  + \mathcal{L} \left(\{\tau_b \} | \bm{\alpha}\right), 
\end{align}
where $\mathcal{L}_0$ comprises the terms that do not depend on $c_{tot}$ or $\bm{\alpha}$, while $\mathcal{L}\left(T_u | c_{tot}\right) $ and $\mathcal{L} \left(\{\tau_b \} | \bm{\alpha}\right)$ are the functions of the total concentration $c_{tot}$ and the ligand concentration ratios $\bm{\alpha}$ denoted here as an ($M \times 1$) vector, respectively. For estimating the individual ligand concentrations, we are only interested in the log-likelihoods that are functions of $c_{tot}$ and $\bm{\alpha}$ and given as
\begin{equation}
\mathcal{L}\left(T_u | c_{tot}\right) = N \ln(c_{tot})-k^+ c_{tot} T_u,
\end{equation}
\begin{equation}\label{l2}
\mathcal{L} \left(\{\tau_b \} | \bm{\alpha}\right) = \sum_{i=1}^{N} \ln \p\left(\tau_{b,i}\right).
\end{equation}
Accordingly, $\mathcal{L}\left(T_u | c_{tot}\right)$ tells us that the total unbound time $T_u$ is informative of the total ligand concentration $c_{tot}$, whereas $\mathcal{L} \left(\{\tau_b \} | \bm{\alpha}\right)$ shows that the individual bound time durations $\{\tau_b \}$ are informative of the ligand concentration ratios $\bm{\alpha}$. Hence, the estimators of individual ligand concentrations introduced in the next section will be based on these two likelihood functions. 

\section{Channel Sensing based on Ligand-Receptor Binding Reaction}
\label{sec:spectrumsensing}
\subsection{System Model}
In the considered scenario, the MC receiver is equipped with only a single type of receptors, whereas there are multiple types of ligands in the channel medium. The following assumptions are made regarding the properties of the receiver and ligands. 
\begin{itemize}
\item There are M different types of ligands in the medium, which have distinct unbinding rates in their reactions with the receptors. We assume that the binding rates of ligand types are equal to $k^+ = 4Da$, with the condition of diffusion-limited propagation, as discussed in Section \ref{sec:statistics}. This assumption is made for the sake of simplicity of the derivations; however, it does not limit the applicability of the estimators to other cases, where binding rates may also differ, as investigated in \cite{singh2017simple} for two types of ligands.   
\item Receptors are assumed to be independent of each other. All receptors are exposed to the same concentration of ligands. In practice, this may correspond to a scenario where the receptors are free to diffuse in a lipid membrane of a cell. We assume that within this membrane the ligands and receptors are homogeneously distributed. 
\item Ligand concentration in the vicinity of receptors is assumed to be stationary during estimation. This assumption is based on the low-pass characteristics of the MC channel, as discussed in the Section \ref{sec:statistics}. We also assume the fluctuations in the concentration of ligands resulting from binding reactions are negligible. 
\item We assume that the unbinding rates of co-existing ligands are known to the receiver. This may correspond to a scenario, where a receiver is hardwired prior to its utilization for the potential set of ligand types that may exist in an application environment. As we will see in Section \ref{sec:absence}, in the case of absence of any ligand type from this set has a slight effect on the overall performance of the estimators. Hence, hardwiring the receiver with a large set of potential ligand types can overcome the limitations of this assumption. 
\end{itemize}

\subsection{Optimal Estimation of Ligand Concentrations and Cram\'er-Rao Lower Bound (CRLB)}
The likelihood function in \eqref{log_like} suggests that we can estimate the concentration of each ligand type by simultaneously inferring the total ligand concentration and concentration ratios of ligand types from the total unbound time $T_u$ and bound time durations $\{\tau_b \}$ of receptors, respectively.

\subsubsection{Optimal Estimation of Total Ligand Concentration}
\label{estimation_total_concentration}
An ML estimator of the total ligand concentration $c_{tot}$ can be found by solving $\partial \mathcal{L}\left(T_u | c_{tot}\right) / \partial c_{tot} = 0$ for $c_{tot}$ that maximizes the likelihood. The resulting ML estimator is obtained as $\hat{c}_{tot} = \frac{N}{k^+ T_u}$. Note that $T_u$, as the sum of $N$ independent exponential random variables ($\tau_{u,i}$'s), is gamma distributed, making its reciprocal $1/T_u$ an inverse gamma-distributed random variable with mean
\begin{equation}\label{Tu}
\E[1/T_u] = \frac{k^+ c_{tot}}{N-1}.
\end{equation}
Hence, the mean of the estimator becomes $\E[\hat{c}_{tot}] = (N/k^+) \times \E[1/T_u] = c_{tot} \times N/(N-1)$, rendering it biased unless $N$ is very large. Therefore, we prefer here using its unbiased version, which is obtained by modifying only the numerator of the biased estimator as follows
\begin{equation}\label{DRUTestimate}
\hat{c}_{tot} = \frac{N-1}{k^+ T_u},
\end{equation}
which is unbiased for $N > 1$ \cite{kuscu2018maximum}. Accordingly, the mean squared error (MSE) of this unbiased estimator is given by its variance \cite{kuscu2018maximum}, i.e.,
\begin{align}
\MSE [\hat{c}_{tot}] = \Var \left[ \hat{c}_{tot}\right] = \frac{c^2_{tot}}{N-2}~~ \text{for} ~~ N > 2. \label{ctot_est_var}
\end{align}
Note that the mean of the total concentration estimator now becomes equal to the actual value of the total ligand concentration, i.e.,
\begin{align}
\E [\hat{c}_{tot}] = c_{tot}. \label{ctot_est_mean}
\end{align}
\subsubsection{Optimal Estimation of Ligand Concentration Ratios}
The ML estimation of the co-existing ligand types' concentration ratios, i.e., $\bm{\hat{\alpha}_{ML}}$, can be performed in the same manner, by solving $\partial \mathcal{L} \left(\{\tau_b \} | \bm{\alpha}\right)  / \partial \alpha_i = 0$ for the $i^\text{th}$ ligand, i.e., 
\begin{equation}\label{l2}
0 =\mathlarger{\sum_{j=1}^{N}} \frac{k_i^- \e^{-\left(k_i^- \tau_{b,j}\right)}} { \sum_{j=1}^M \alpha_j k_j^- \e^{-k_j^- \tau_{b,i}}}~~~ \text{for}~ i \in \{1, \dots,	 M \}.
\end{equation}
The expression in \eqref{l2} does not lend itself to an analytical solution for ML estimate $\bm{\hat{\alpha}_{ML}}$, and necessitates numerical approaches. In fact, the problem of Bayesian inference from mixture of exponential distributions is generally tackled by computationally expensive iterative algorithms, e.g., expectation-maximization (EM) algorithm \cite{hasselblad1969estimation, jewell1982mixtures}, which cannot be considered feasible for resource-limited bionanomachines, and thus they are disregarded in this study. 
\subsubsection{Optimal Estimation of Individual Ligand Concentrations} 
The optimal ML estimator of the concentration of individual ligand types can be given as the product of the estimators of the total ligand concentration and ligand concentration ratios, i.e., 
\begin{equation}\label{cML}
\bm{\hat{c}_{ML}} = \hat{c}_{tot} \bm{\hat{\alpha}_{ML}}. 
\end{equation}
The derivation of the variance for this optimal ML estimator is not analytically tractable. Instead, in the following section, we derive a lower bound on its variance through Cram\'er-Rao formalism.
\subsubsection{Error Bound}
The Cram\'er-Rao lower bound (CRLB) gives the minimum variance of any unbiased estimator as the inverse of the Fisher information \cite{kay1993fundamentals}. The ML estimator of the total ligand concentration, whose variance is given in \eqref{ctot_est_var}, already achieves the CRLB \cite{kuscu2018maximum}. To obtain the CRLB for the estimator of individual ligand concentrations, we first need to derive the bound for the ML estimator of ligand concentration ratios.  The Fisher information of the concentration ratio vector $\bm{\alpha}$ is an $(M \times M)$ matrix, which is given by the negative expectation of the Hessian matrix, i.e.,  
\begin{align}\label{fisher}
\bm{I_\alpha}= -\bm{\E}[\bm{H_\alpha}]. 
\end{align}
The elements of the Hessian matrix are given by the second-order partial derivatives of the log-likelihood function that governs the relation between bound time durations and ligand concentration ratios, i.e., 
\begin{align}  \label{hessian}
\bm{H_\alpha}(i,j) & = \frac{\partial^2}{\partial \alpha_i \partial \alpha_j} \mathcal{L} \left(\{\tau_b \} | \bm{\alpha}\right) \\ \nonumber
& = - \sum_{l=1}^N \frac{k_i^- k_j^-}{\p\left(\tau_{b,l}\right)^2}  \e^{-(k_i^- + k_j^-) \tau_{b,l}}.
\end{align}
Substituting \eqref{hessian} in \eqref{fisher}, the elements of the Fisher information matrix are obtained as follows
\begin{align}\label{fisher2}
\bm{I_\alpha}(i,j) & = -\E \left[ - \sum_{l=1}^N \frac{1}{\p\left(\tau_{b,l}\right)^2} k_i^- k_j^- \e^{-(k_i^- + k_j^-) \tau_{b,l}}\right] \\ \nonumber
& = \sum_{l=1}^N \E \left[  \frac{1}{\p\left(\tau_{b,l}\right)^2} k_i^- k_j^- \e^{-(k_i^- + k_j^-) \tau_{b,l}} \right] \\ \nonumber
& = N k_i^- k_j^- \int_{0}^{\infty} \frac{1}{\p\left(\tau_b'\right)}  \e^{-(k_i^- + k_j^-) \tau_b'} d\tau_b'. 
\end{align}

The CRLB is then given by the inverse of the $i^\text{th}$ diagonal element of the inverse Fisher information matrix for the estimation of concentration ratio of the  $i^\text{th}$ ligand, i.e.,
\begin{align}\label{eq:CRLBalpha}
\Var[\hat{\alpha}_{ML, i}]_{LB} = \bm{I_\alpha^{-1}}(i,i).
\end{align}

Given that the ML estimator of total ligand concentration $\hat{c}_{tot}$ already achieves the lower bound, the CRLB for the concentration estimator given in \eqref{cML} can be written as 
\begin{align} \label{eq:totalvarianceLB}
&\bm{\Var[\hat{c}_{ML}]_{LB}} = \Var[\hat{c}_{tot}] \bm{\Var[\hat{\alpha}_{ML}]_{LB}} + \\ \nonumber
& \Var[\hat{c}_{tot}] \left( \bm{\E[\hat{\alpha}_{ML}]} \odot \bm{\E[\hat{\alpha}_{ML}]}\right) + \bm{\Var[\hat{\alpha}_{ML}]_{LB}} \E[\hat{c}_{tot}]^2, 
\end{align}
where $\bm{\E[\hat{\alpha}_{ML}]} = \bm{\alpha}$; $\bm{\Var[\hat{\alpha}_{ML}]_{LB}}$ is an $(M \times 1)$ vector with the $i^\text{th}$ element given by \eqref{eq:CRLBalpha}, and $\odot$ denotes the Hadamard product, i.e., $(\bm{A} \odot \bm{B})_{i,j} = (\bm{A})_{i,j} (\bm{B})_{i,j}$. Note that since the optimal ML estimator achieving the CRLB is an unbiased estimator, the lower bound on the MSE is equal to the CRLB on the variance, i.e., $\bm{\MSE[\hat{c}_{ML}]_{LB}} = \bm{\Var[\hat{c}_{ML}]_{LB}}$. 

\subsection{Suboptimal Estimation of Ligand Concentrations}
The optimal ML estimation of the concentration ratios, and thus the individual concentrations of ligands, is not feasible for resource-limited bio-nanomachines, as it requires complex numerical calculations. To overcome this problem, we propose a novel practical method based on method of moments (MoM) \cite{kay1993fundamentals} to estimate the concentration ratios of ligand types. The method relies on statistically binning the receptor bound times into a number of time intervals determined by the unbinding rates of existing ligands, instead of using the exact bound time durations. Next, we investigate two versions of this estimation method, which provide unbiased and simplified biased estimation of the concentration of each ligand type. 
\begin{figure}[!t]
	\centering
	\includegraphics[width=9cm]{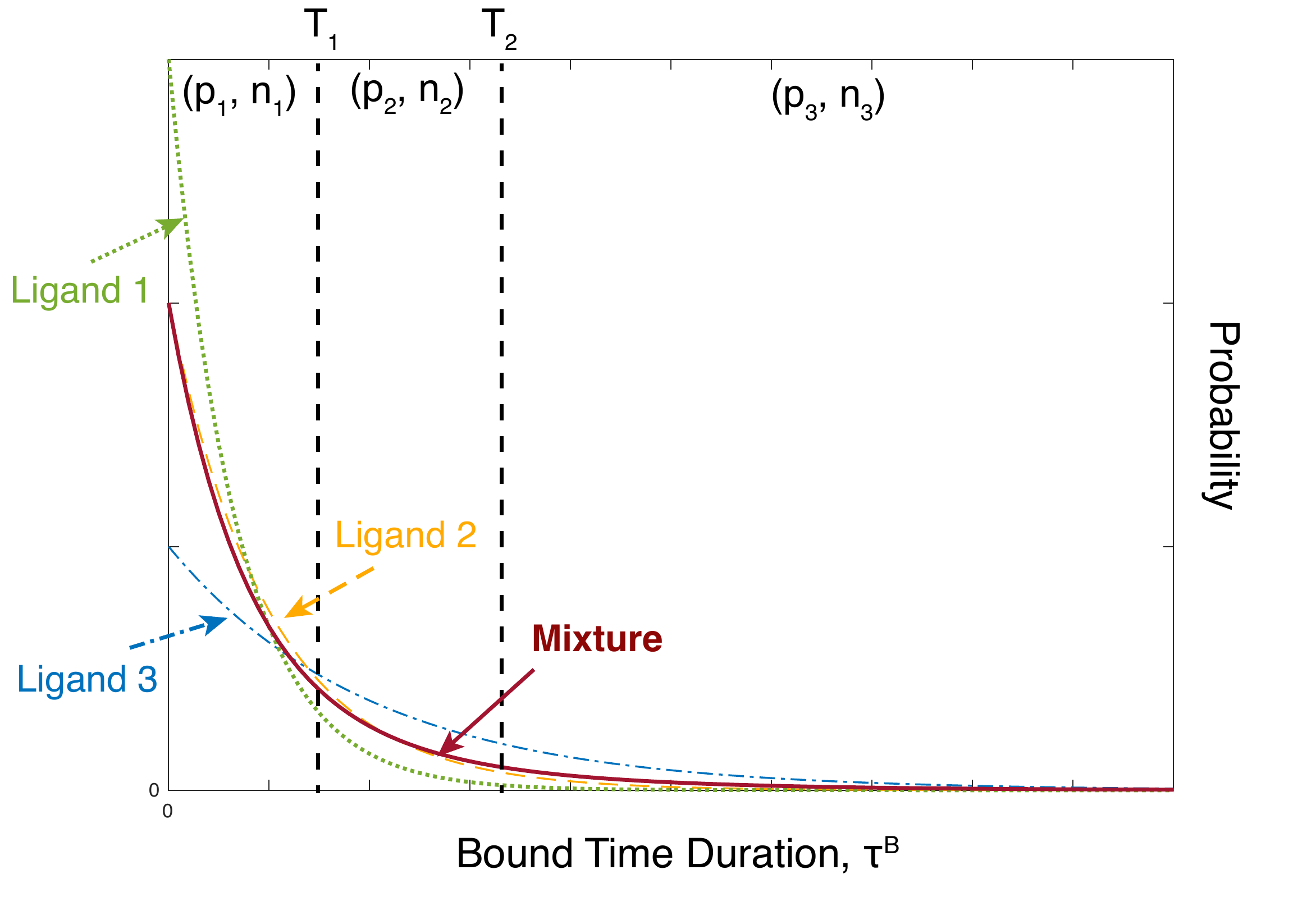}
	\caption{Probability distribution of receptor bound time duration for a mixture of $3$ ligands, which is a mixture of exponential distributions. The regions separated by the time thresholds ($T_1$ and $T_2$) are marked with the corresponding probability of observing a binding time duration and the number of binding events in those regions, i.e., $p_i, n_i$, respectively.}
	\label{fig:mixture}
\end{figure}

\subsubsection{Unbiased Estimation of Ligand Concentrations}
\label{sec:unbiased}
The proposed estimation method for ligand concentration ratios is based on counting the number of binding events that fall in specific time intervals. The number of the time intervals is equal to the number of ligand types, as demonstrated in Fig. \ref{fig:mixture}. These non-overlapping intervals are defined by time thresholds, which can be taken as proportional to the inverse of the unbinding rates of ligands, i.e., 
\begin{equation} \label{timethreshold}
T_i = \nu/k_i^- ~~\text{for}~~i \in \{1, \dots M-1 \}, 
\end{equation}
given that all ligand types are sorted in decreasing order of unbinding rate, i.e., increasing order of their affinity with the receptors. Here, $\nu > 0$ is a proportionality constant, which will be optimized in Section \ref{sec:performance}. Later, we will also show that this transduction scheme is suitable for biological MC devices, as it can be implemented by active receptors based on well-known KPR scheme. 

The probability of observing a ligand binding event of a duration that falls in a time interval between two time thresholds can be written as
\begin{equation}
\p_l = \int_{T_{l-1}}^{T_l} \p(\tau_b') d\tau_b' =  \sum_{i=1}^M \alpha_i \left( \e^{-(k_i^- T_{l-1})} - \e^{-(k_i^- T_{l})} \right),
\end{equation}
where we set $T_0 = 0$ and $T_M = +\infty$. In matrix notation, the probabilities can be written as
\begin{equation} \label{eq:probmatrix}
\bm{\p} = \bm{S} \bm{\alpha},
\end{equation}
where $\bm{\p}$ is an ($M\times1$) probability vector with the elements $\p_l$, and $\bm{S}$ is an ($M \times M$) matrix with the elements 
\begin{equation} \label{eq:Smatrix}
s_{i,j} = \e^{-(k^-_j T_{i-1})} - \e^{-(k^-_j T_{i})}.
\end{equation}

The number of binding events that fall in each interval follows binomial distribution with the mean and the variance given by
\begin{equation}
\bm{\E[n]} = \bm{\p} N,
\label{eq:meann}
\end{equation}
\begin{equation}
\bm{\Var[n]} = \left( \bm{\p} \odot (1-\bm{\p}) \right) N,
\label{eq:varn}
\end{equation}
where $\bm{n}$ is an $(M \times 1)$ vector with the vector elements $n_i$ being the number of binding events whose durations are within the $i^\text{th}$ time interval defined by $T_{i-1}$ and $T_i$.

We can now apply the MoM for the estimation of ligand concentration ratios by employing only the first moment. In other words, we match the expected number of binding events in a time interval to the actual number of binding events observed for the same interval, i.e.,
\begin{align} \label{mom1}
\bm{n} = \bm{\hat{\p}} N = \bm{S} \bm{\hat{\alpha}}N,  
\end{align}
where the hat denotes the estimated parameters. Rearranging the parameters in \eqref{mom1}, we obtain an estimator for the ligand concentration ratio vector:
\begin{align} \label{Wmatrix}
\bm{\hat{\alpha}} = \left(\frac{1}{N}\right) \bm{W} \bm{n},  
\end{align}
where $\bm{W} = \bm{S}^{-1}$, i.e., the inverse of $\bm{S}$ matrix, which is also an ($M \times M$) matrix with elements $w_{i,j}$. Note that the estimated concentration ratio of each ligand type, i.e.,
\begin{equation}
\hat{\alpha}_l = \left(\frac{1}{N}\right)  \sum_{i=1}^M n_i w_{l,i}, \label{ratio_est}
\end{equation}
becomes the weighted sums of $M$ correlated binomial random variables with the weights $w_{l,i}$. 

Combining the ratio estimator with the unbiased estimator of total ligand concentration introduced in Section \ref{estimation_total_concentration}, we can obtain an estimator for the concentration of each ligand type as follows 
\begin{align}
\bm{\hat{c}} = \hat{c}_{tot} \bm{\hat{\alpha}},
\end{align}
with
\begin{align}
\hat{c_l} &= \frac{N-1}{N} \frac{1}{k^+ T_u}  \sum_{i=1}^M n_i w_{l,i}, \\ \nonumber
&\approx \frac{1}{k^+ T_u}  \sum_{i=1}^M n_i w_{l,i}, ~~~ \text{for}~ N \gg 1.
\end{align}
The variance of this estimator can be calculated as follows
\begin{align} \label{eq:totalvarianceunbiased}
\bm{\Var[\hat{c}]} &= \Var[\hat{c}_{tot}] \bm{\Var[\hat{\alpha}]} + \Var[\hat{c}_{tot}] \left( \bm{\E[\hat{\alpha}]} \odot \bm{\E[\hat{\alpha}]}\right) \\ \nonumber
&+ \bm{\Var[\hat{\alpha}]} \E[\hat{c}_{tot}]^2, 
\end{align}
where the variance and the mean of the unbiased estimator of total concentration $\hat{c}_{tot}$ are given in \eqref{ctot_est_var} and \eqref{ctot_est_mean}, respectively. On the other hand, the variance of the ratio estimator given in \eqref{ratio_est} can be written for the $l^\text{th}$ ligand as
\begin{equation}
\Var[\hat{\alpha}_l]= \frac{1}{N^2}  \sum_{i=1}^M \sum_{j=1}^M w_{l,i} w_{l,j} ~\Cov[n_i, n_j],  \label{eq:est_variance}
\end{equation}
with the covariance function
\begin{equation}
\Cov[n_i, n_j]  = 
\begin{cases}
\Var[n_i],  & \text{if } i = j,\\
-p_i p_j N,    & \text{otherwise}.
\end{cases}
\label{eq:est_covariance}
\end{equation}
The expected value of the ratio estimator is equal to the actual value of the concentration ratio vector $\bm{\alpha}$, i.e.,
\begin{align}
\bm{\E[\hat{\alpha}]} = \left(\frac{1}{N}\right) \bm{W} \bm{\E[n]} = \bm{W} \bm{\p} = \bm{S^{-1}} \bm{\p} = \bm{\alpha},
\end{align}
validating the unbiasedness of the overall concentration estimator, i.e., 
\begin{align}
	\bm{\E[\hat{c}]} = \E[\hat{c}_{tot}] \bm{\E[\hat{\alpha}]} = c_{tot} \bm{\alpha} = \bm{c}.
\end{align}
Therefore, resulting MSE for this unbiased concentration estimator becomes equal to its variance, i.e.,
\begin{align} \label{eq:MSEunbiased}
\bm{\MSE[\hat{c}]} &=  \bm{\Var[\hat{c}]}.
\end{align}

\subsubsection{Biased Estimation of Ligand Concentrations}
\label{sec:biased}
We also introduce a biased version of the concentration ratio estimator, which has a simplified design, enabled when the time thresholds given in \eqref{timethreshold} are set sufficiently large. In this case, we can neglect the noisy contributions of the ligand types that have higher unbinding rates than the ligand type that is being estimated. When the thresholds are much larger than the corresponding unbinding rates, i.e., $T_i \gg 1/k_i^-$, $\bm{S}$ matrix, whose elements are given in \eqref{eq:Smatrix}, can be approximated by an upper triangular matrix, i.e., 
\begin{equation}\label{eq:Hmatrix}
\bm{H} = \bm{S}|_{T_i \gg 1/k_i^-},
\end{equation}
with the matrix elements given as 
\begin{equation}
h_{i,j} = 
\begin{cases}
s_{i,j}, & \text{if } i < j,\\
\e^{-(k^-_i T_{i-1})} ,& \text{if } i = j,\\
0,              & \text{otherwise}.
\end{cases}
\label{eq:Hmatrix2}
\end{equation}
This approximation results in the following ratio estimator,
\begin{equation}
\bm{\hat{\alpha}^\ast} = \left(\frac{1}{N}\right) \bm{R} \bm{n},
\end{equation}
where $\bm{R} = \bm{H}^{-1}$, is also an upper triangular matrix. The elements of $\bm{R}$ can be recursively calculated as follows
\begin{equation}
r_{i, j} = \kappa_j \left( \mathds{1}_{\{i=j\}} - \sum_{\gamma=1}^{j-i} r_{i+\gamma, j} ~\theta_{i+\gamma, i}  \right),
\end{equation}
where $\mathds{1}_{\{i=j\}}$ is the indicator function which is equal to $1$ if $i=j$, and $0$ otherwise; $\kappa_j = \e^{k_j^- T_{j-1}}$, and $\theta_{i, j} = \e^{-(k_i^- T_{i-j-1})} - \e^{-(k_i^- T_{i-j})}$. 
Since $\bm{R}$ is an upper triangular matrix, the estimator for the concentration ratio of the $l^\text{th}$ ligand type can be written as the sum of $M-l+1$ terms, i.e., 
\begin{equation}
\hat{\alpha}^\ast_l = \frac{1}{N} \sum_{i=0}^{M-l} n_{M-i} r_{l, M-i}.
\end{equation}
This substantially simplifies the ratio estimation of the ligand types with the highest affinities, which, in most cases, are the most relevant ligands for information transfer in MC. Similar to \eqref{eq:est_variance}, the variance of this estimator can be written as
\begin{equation}
\Var[\hat{\alpha}^\ast_l]=   \frac{1}{N^2} \sum_{i=0}^{M-l} \sum_{j=0}^{M-l}  r_{l,M-i} r_{l,M-j} \Cov[n_{M-i}, n_{M-j}],
\end{equation}
where $\Cov[n_{M-i}, n_{M-j}]$ can be calculated using \eqref{eq:est_covariance}. 
The mean of this estimator is given as
\begin{equation}
\bm{\E[\hat{\alpha}^\ast]}=  \left( \frac{1}{N}\right) \bm{R} \bm{\E[n]} = \bm{R} \bm{p}. \label{eq:simplifiedmean}
\end{equation}
As is clear from \eqref{eq:simplifiedmean}, this is a biased estimator, due to the residuals resulting from the approximation of the $\bm{S}$ matrix with an upper triangular matrix $\bm{H}$ given in \eqref{eq:Hmatrix}. The resulting bias can be calculated as the difference between expected value of the estimation and the actual value of the concentration ratios, i.e.,
\begin{align}
\bm{\Delta[\hat{\alpha}^\ast]} &= \bm{\E[\hat{\alpha}^\ast]} - \bm{\alpha} \\ \nonumber
&= \bm{R} \bm{p} - \bm{S}^{-1} \bm{p} \\ \nonumber
&= \left(\bm{R} - \bm{W}\right) \bm{p}.
\end{align}
The MSE of this biased ratio estimator can then be written as
\begin{equation}
\bm{\MSE[\hat{\alpha}^\ast]}=  \bm{\Var[\hat{\alpha}^\ast]} + \left( \bm{\Delta[\hat{\alpha}^\ast]} \odot \bm{\Delta[\hat{\alpha}^\ast]} \right). 
\end{equation}
The resulting biased estimator of the concentrations of individual ligand types can be given as 
\begin{align}
\bm{\hat{c}^\ast} &= \hat{c}_{tot} \bm{\hat{\alpha}^\ast},
\end{align}
with the matrix elements calculated as	
\begin{align}
\hat{c_l}^\ast &= \frac{N-1}{N} \frac{1}{k^+ T_u}  \sum_{i=0}^{M-l} n_{M-i} r_{l, M-i}, \\ \nonumber
&\approx \frac{1}{k^+ T_u}  \sum_{i=0}^{M-l} n_{M-i} r_{l, M-i}, ~~~ \text{for}~ N \gg 1,
\end{align}
and the corresponding variance is obtained via
\begin{align}
\bm{\Var[\hat{c}^\ast]} &= \Var[\hat{c}_{tot}] \bm{\Var[\hat{\alpha}^\ast]} \\ \nonumber
&+ \Var[\hat{c}_{tot}] \left( \bm{\E[\hat{\alpha}^\ast]} \odot \bm{\E[\hat{\alpha}^\ast]}\right) + \bm{\Var[\hat{\alpha}^\ast]} \E[\hat{c}_{tot}]^2.
\end{align}
The bias of this estimator is given as
\begin{align}
\bm{\Delta[\hat{c}^\ast]} &= \bm{\E[\hat{c}^\ast]} - \bm{c} \\ \nonumber
&= c_{tot} \left( \bm{\E[\hat{\alpha}^\ast]} - \bm{\alpha} \right) =  c_{tot} \bm{\Delta[\hat{\alpha}^\ast]}.
\end{align}
Finally, the resulting MSE can be obtained as follows
\begin{align} \label{eq:MSEbiased}
\bm{\MSE[\hat{c}^\ast]} =  \bm{\Var[\hat{c}^\ast]} + \left( \bm{\Delta[\hat{c}^\ast]} \odot \bm{\Delta[\hat{c}^\ast]} \right).
\end{align}

\section{Performance Analysis}
\label{sec:performance}
We evaluate the performance of the proposed channel sensing methods in terms of normalized MSE (NMSE) averaged over all ligands co-existing in the mixture, i.e., 
\begin{align}
\langle \NMSE[\bm{\hat{c}]} \rangle = \frac{1}{M} \sum_{i=1}^M \NMSE[\hat{c}_i] = \frac{1}{M} \sum_{i=1}^M \frac{\MSE[\hat{c}_i]}{c_i^2}.
\end{align}
The average NMSE can be calculated for the optimal ML estimator $\bm{\hat{c}_{ML}}$ and the simplified estimator $\bm{\hat{c}^\ast}$ in the same way. Note that with the normalization, we render the analysis independent of the total ligand concentration $c_{tot}$. Hence, the performance of the proposed method in terms of the normalized performance metric only depends on the number of unbound and bound time duration samples, relative affinities of the ligand types with the receptors, number of ligand types, and the ligand concentration ratios. 

For the simplicity of the analysis, we assume, without loss of generality, that the unbinding rates of ligand types are indexed in decreasing order, i.e., $k^-_1 > k^-_2 > ... > k^-_M$, and we define the following rule to describe the relation between them:
\begin{align}
k^-_{M-i} = \chi^i k^-_M ~~~ \text{for}~ i \in \{1, \dots, M-1 \},
\end{align}
where $\chi$ is the similarity parameter that provides a measure of pairwise similarity between different ligand types in the mixture. Its effect on the performance of the estimators will be discussed in Section \ref{sec:similarity}. In the rest of the analysis, we set $\chi = 5$, such that the ratio of the unbinding rates between the two most similar ligands is $\chi = 5$. We also set $k_M^- = 1\text{s}^{-1}$. 

The time thresholds, defined in \eqref{timethreshold}, are taken as proportional to the inverse of the unbinding rates of the corresponding ligand types, i.e., $T_i = \nu/k_i^-$, where $\nu$ is the proportionality constant. For each system setting we analyze with sub-optimal unbiased estimator, we also perform the optimization of $\nu$ for the minimum average NMSE, i.e., 
\begin{align}
	\nu_{opt} = \argmin_{\nu > 0} \langle \NMSE[\bm{\hat{c}}] \rangle ,
\end{align}
 and provide the $\nu$-optimized value of the average NMSE together with the performance of the suboptimal unbiased and simplified biased estimators, and with the corresponding CRLB. The obtained values of $\nu_{opt}$ are different for each setting; however, we find that they concentrate around $\nu_{opt} = 3$ (data not shown). Therefore, in the performance evaluation of the sub-optimal unbiased estimator and in calculating the CRLB, we set $\nu = 3$. For the simplified biased estimator, however, the value of $\nu$ is constrained by the fact that the simplification is based on the assumption that $T_i \gg 1/k_i^-$. With this condition, we obtain an upper triangular estimator matrix, which, in turn, simplifies the calculations required for estimation (see  \eqref{eq:Hmatrix} and \eqref{eq:Hmatrix2}). In our analysis, we conclude that setting $\nu = 5$ is sufficient for the validity of this assumption. Moreover, throughout the analysis we set the default number of samples and the default number of ligand types in the channel as $N = 10000$ and $M = 5$, respectively. 

Given the default system setting above, next we evaluate the sensing performance for varying number, similarity, and ratio distribution of ligand types, and varying number of samples. We will also evaluate the sensing performance in two particular cases, where some of the ligand types considered in the estimator do not actually exist in the channel medium, and new types of ligands that are not hardwired into the estimators are added to the medium. We provide a brief comparison of the investigated channel sensing methods in terms of their requirements, properties and performance in Table \ref{table:comparison}.

\begin{table*}[!t]\scriptsize
	\centering
	\begin{threeparttable}
		\centering
		\caption{Comparison of Channel Sensing Methods}
		\label{table:comparison}
		\begin{tabular}{llllll}
			\toprule	
			\textbf{}   & \textbf{Optimal} & \textbf{Suboptimal} & \textbf{Suboptimal Simplified}&  \textbf{Suboptimal $\nu$-Optimized} \\ 
			\textbf{}   & \textbf{Estimation $\bm{\hat{c}_{ML}}$} & \textbf{Estimation $\bm{\hat{c}}$} & \textbf{Estimation  $\bm{\hat{c}^\ast}$}&  \textbf{Estimation  $\bm{\hat{c}}_\nu$} \\
			\toprule
			\textbf{Method}  & ML & ML + MoM  & ML + MoM &  ML + MoM  \\ \midrule
			\textbf{Required Statistics}   & $T_u$, $\{\tau_b\}_N$ & $T_u$, $n_i$ for $i \in \{1, \dots, M\}$ & $T_u$, $n_i$ for $i \in \{1, \dots, M\}$ & $T_u$, $n_i$ for $i \in \{1, \dots, M\}$ \\ \midrule
			\textbf{Biasedness}         & Unbiased      & Unbiased  & Biased    &  Unbiased    \\ \midrule
			\textbf{Computational} &  High, requires   & Medium, requires        & Low, requires  &    Medium, requires \\
			 \textbf{Complexity} 					 &  iterative methods    & weighted sum of M terms & weighted sum of (M-i+1) terms& weighted sum of M terms\\
			 					 &  & and division, for each ligand type  & and division, for $i^\text{th}$ ligand type  & and division, for each ligand type \\ \midrule
			\textbf{MSE}   & Low \eqref{eq:totalvarianceLB}  & Medium \eqref{eq:MSEunbiased}  & High \eqref{eq:MSEbiased}         &  Medium \eqref{eq:MSEunbiased}   \\
			\bottomrule
		\end{tabular}%
	\end{threeparttable}	
\end{table*}%

\subsection{Effect of Number of Ligand Types in the Mixture}
\label{sec:numberligandtype}

\begin{figure}[!t]
	\centering
	\includegraphics[width=9cm]{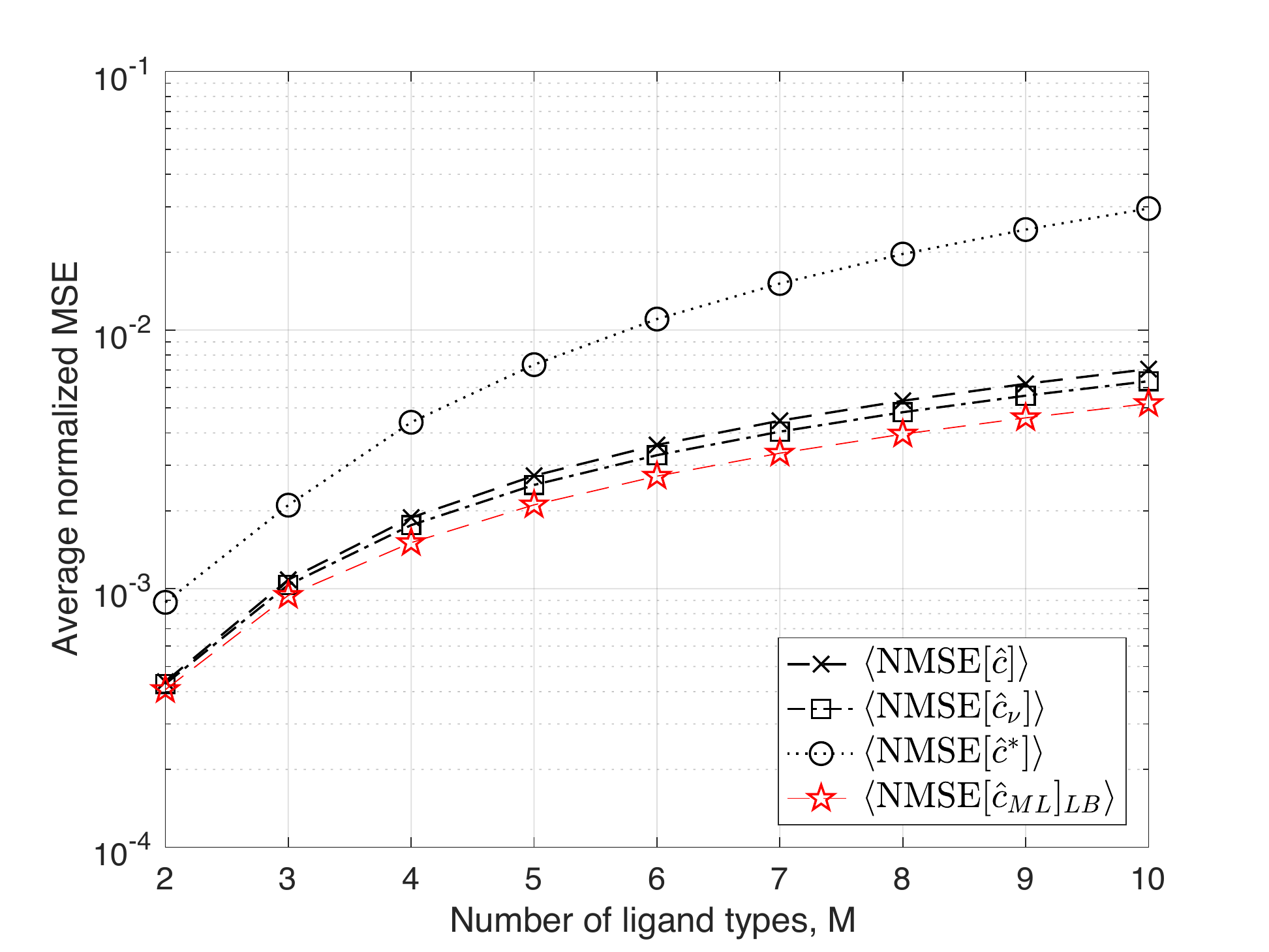}
	\caption{Average NMSE with varying number of ligand types, $M$, for optimal $\bm{\hat{c}_{ML}}$, suboptimal unbiased $\bm{\hat{c}}$, suboptimal biased $\bm{\hat{c}^\ast}$, and $\nu$-optimized unbiased $\bm{\hat{c}_{\nu}}$ estimators.}
	\label{fig:numberligandtype}
\end{figure}

The first analysis is carried out for varying number of ligand types $M$. This is a critical parameter that depends on the interference characteristics of the MC channel and the utilized multiple access scheme. The results are provided in Fig. \ref{fig:numberligandtype}, for CRLB, sub-optimal unbiased estimator, $\nu$-optimized unbiased estimator, and simplified biased estimator. We assume that the concentration ratios of ligand types are equal in all cases, i.e., $\alpha_i = 1/M$ for $i \in \{1, \dots, M\}$.  As expected, the NMSE is increasing with increasing $M$; however, the channel sensing methods demonstrate acceptable performance even when the channel is crowded by $10$ different types of ligands. The results also reveal that the unbiased estimator with $\nu = 3$ and the $\nu$-optimized estimator almost achieve the CRLB, especially when $M<4$, hence, they can be considered highly efficient. The performance of the simplified estimator follows the same trend; however, the resulting error is almost an order of magnitude larger than the unbiased estimators when $M$ is high.

\begin{figure}[!t]
	\centering
	\includegraphics[width=9cm]{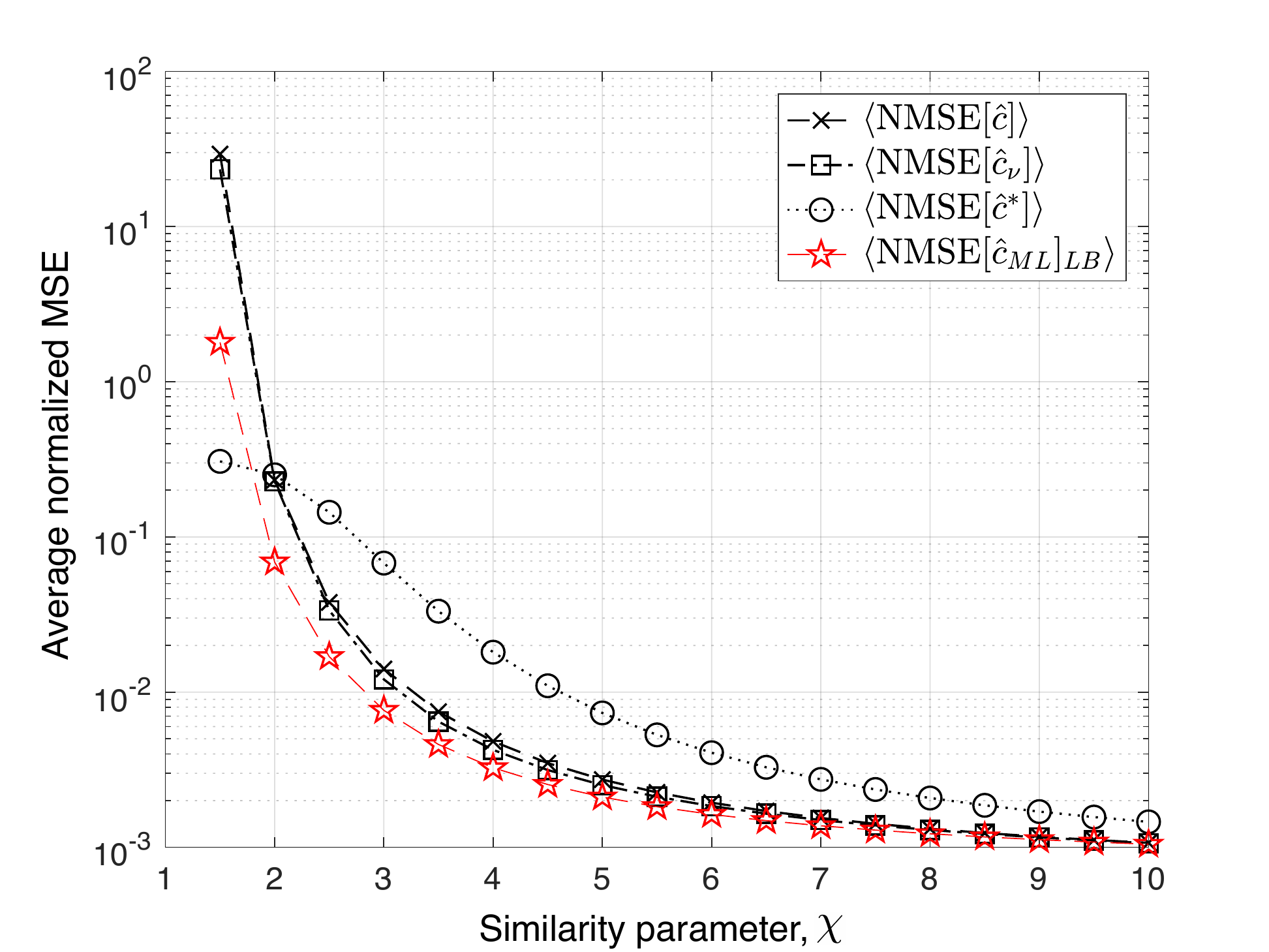}
	\caption{Average NMSE with varying similarity between ligand types, $\chi$, for optimal $\bm{\hat{c}_{ML}}$, suboptimal unbiased $\bm{\hat{c}}$, suboptimal biased $\bm{\hat{c}^\ast}$, and $\nu$-optimized unbiased $\bm{\hat{c}_{\nu}}$ estimators.}
	\label{fig:similarity}
\end{figure}

\subsection{Effect of Similarity between Ligand Types}
\label{sec:similarity}
The similarity of the ligands co-existing in the channel has substantial impact on the performance of the channel sensing methods, as demonstrated in Fig. \ref{fig:similarity}. An increase in the similarity, reflected by a decreasing $\chi$, reduces the capability of the sensing method to discriminate between different ligand types from the bound time duration data. The results indicate that it is not possible to accurately sense the channel with the unbiased estimators when $\chi < 2$ and $M \ge 5$. Interestingly, however, the simplified biased estimator provides superior performance in this range of similarity, even compared to the CRLB. This implies that neglecting the stochastic contribution of the ligands with lower affinities results in better error performance, when the ligands manifest very similar affinities with the receptors.   

\subsection{Effect of Number of Unbound/Bound Duration Samples}
\label{sec:numberofsamples}
\begin{figure}[!t]
	\centering
	\includegraphics[width=9cm]{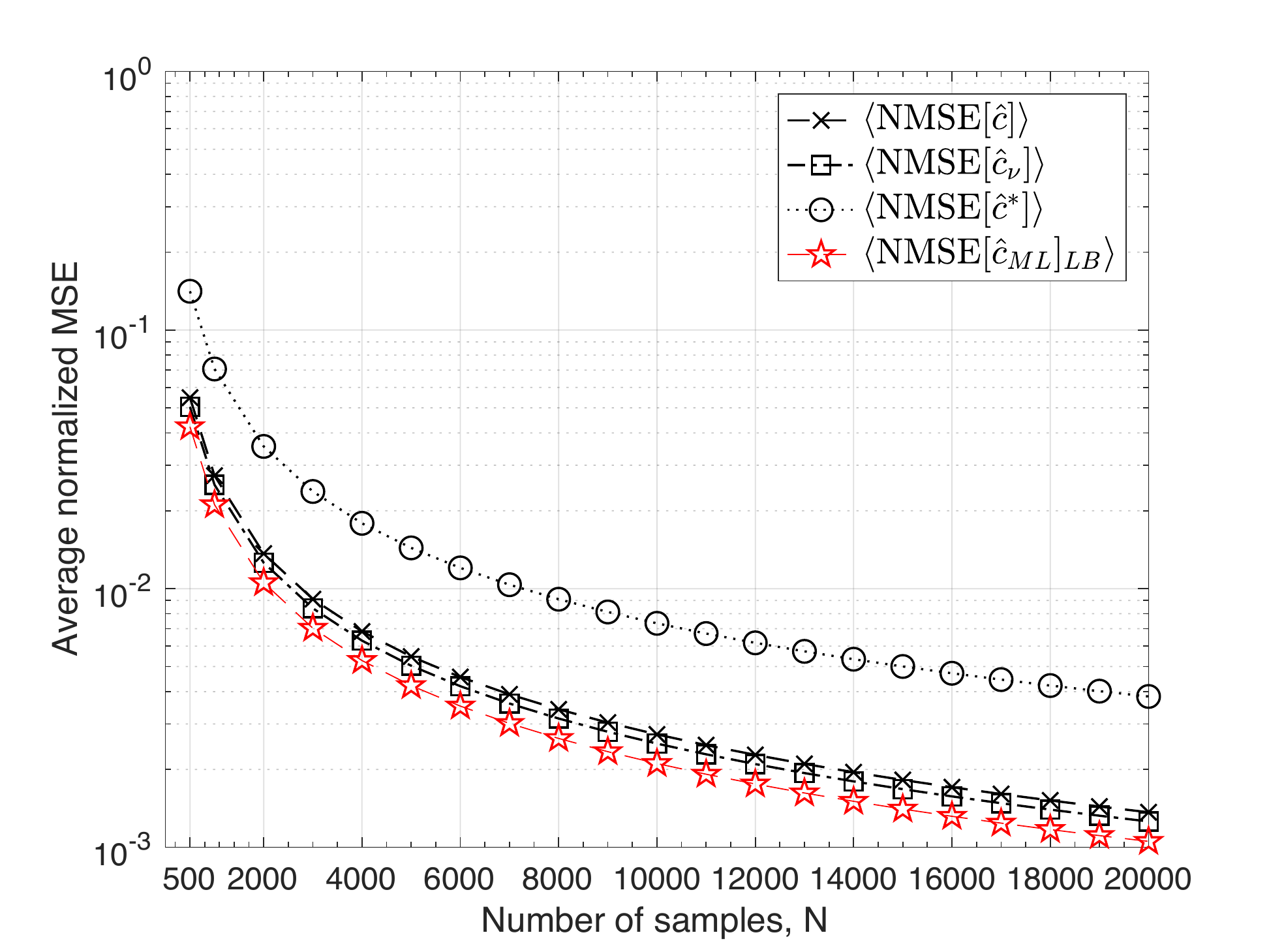}
	\caption{Average NMSE with varying number of unbound and bound time duration samples, $N$, for optimal $\bm{\hat{c}_{ML}}$, suboptimal unbiased $\bm{\hat{c}}$, suboptimal biased $\bm{\hat{c}^\ast}$, and $\nu$-optimized unbiased $\bm{\hat{c}_{\nu}}$ estimators.}
	\label{fig:numberofsamples}
\end{figure}

The number of samples affects the performances of both the ratio estimator and the total concentration estimator $\hat{c}_{tot}$. As a result, the overall impact on the estimation of individual concentrations by all types of estimators is remarkable, as demonstrated in Fig. \ref{fig:numberofsamples}. The relation between the average NMSE and the number of samples follows the same trend for all estimators, and the unbiased estimator has acceptable accuracy even when the number of samples $N = 500$, and $M = 5$. Note that the unbiased estimators, $\bm{\hat{c}}$ and $\bm{\hat{c}_\nu}$, are highly efficient as they perform very closely to the CRLB.

\subsection{Effect of Concentration Ratios of Ligands}
\label{sec:weight}
\begin{figure*}[!t]
	\centering
	\subfigure[]{
		\includegraphics[width=8cm]{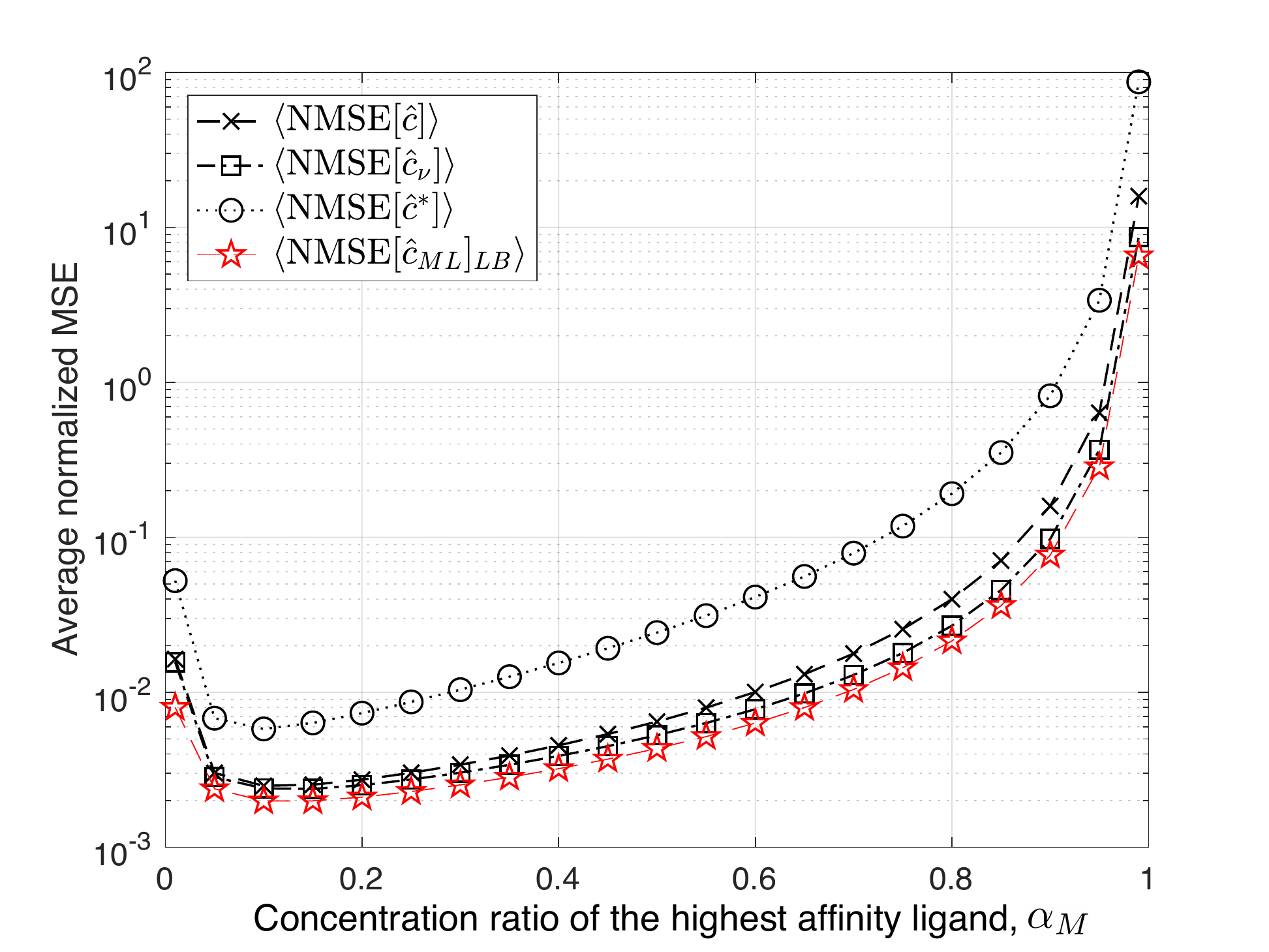}
		\label{fig:weightavg}
	}
	\subfigure[]{
		\includegraphics[width=8cm]{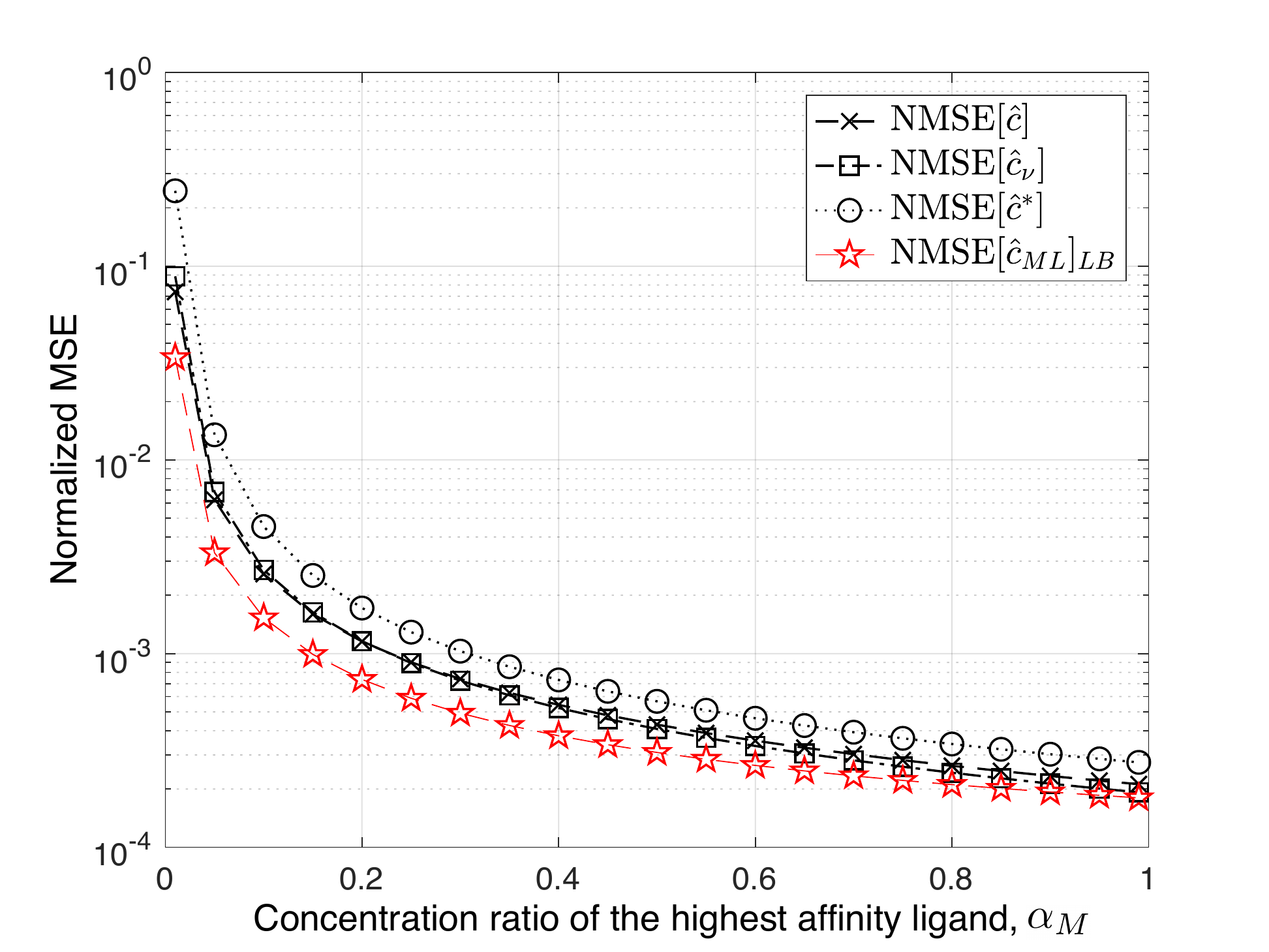}
		\label{fig:weightonly}
	}
	\caption{(a) Average NMSE with varying concentration ratio of the highest-affinity ligand type, $\alpha_M$, for optimal $\bm{\hat{c}_{ML}}$, suboptimal unbiased $\bm{\hat{c}}$, suboptimal biased $\bm{\hat{c}^\ast}$, and $\nu$-optimized unbiased $\bm{\hat{c}_{\nu}}$ estimators. (b) NMSE in the estimation of concentration of the highest-affinity ligand type, $c_M$, with varying concentration ratio of the highest-affinity ligand type, $\alpha_M$, for optimal $\bm{\hat{c}_{ML}}$, suboptimal unbiased $\bm{\hat{c}}$, suboptimal biased $\bm{\hat{c}^\ast}$, and $\nu$-optimized unbiased $\bm{\hat{c}_{\nu}}$ estimators.}
	\label{fig:weight}
\end{figure*}

Lastly, we evaluate the sensing performance for the case of heterogeneous distribution of concentration ratios, i.e., $\alpha_i = c_i/c_{tot}$. In particular, we change the concentration ratio of the ligand type that has the highest affinity with the receptors, i.e., $\alpha_M$, while keeping the ratios of the other ligand types homogeneously distributed, i.e., 
\begin{align}
\alpha_i = \frac{1-\alpha_M}{M-1} ~~~\text{for}~~ i \in \{1, 2, \cdots M-1 \}.
\end{align}
The results are provided in terms of NMSE averaged over all ligand types in Fig. \ref{fig:weightavg}, and in terms of NMSE of concentration estimation of only the highest-affinity ligand in Fig. \ref{fig:weightonly}. Given that there are $M = 5$ different types of ligands in the channel, the average NMSE is minimized when the weights are almost uniformly distributed, i.e., when $\alpha_M \approx 0.2$. Interpreting both results together, we see that while the accuracy of the concentration estimation for the highest-affinity ligand, i.e., $c_M$, significantly increases for very high values of $\alpha_M$, the overall performance of the channel sensing deteriorates. In an MC application, we can expect that the molecules of interest, i.e., information-carrying molecules, would be the ligands that have the highest affinity with the employed receptors. Hence, the results show that the proposed channel sensing methods can be effectively used to eliminate the interference of lower-affinity ligands for improving the detection performance.  

\begin{figure*}[!t]
	\centering
	\subfigure[]{
		\includegraphics[width=8cm]{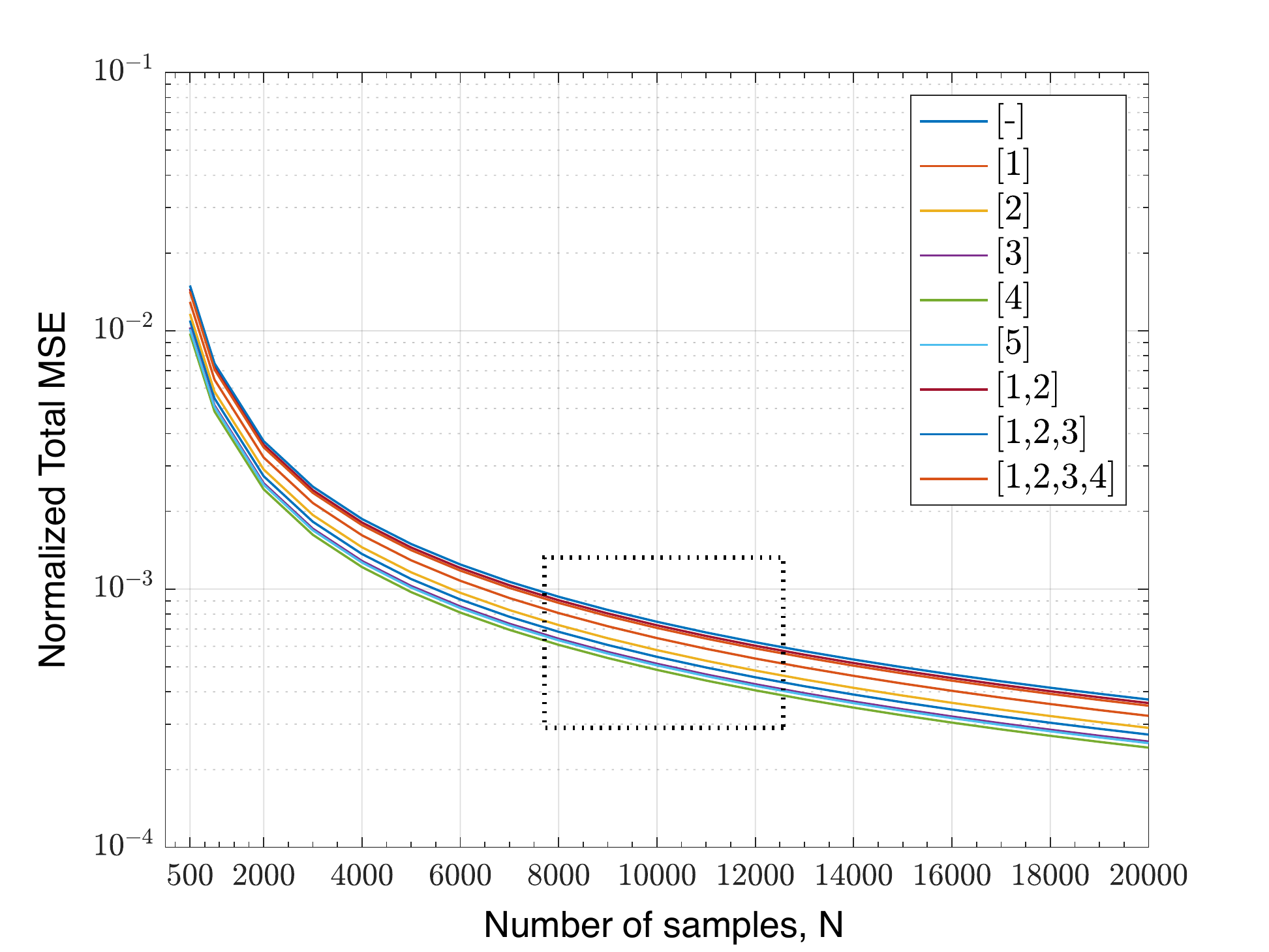}
		\label{fig:absence_a}
	}
	\subfigure[]{
		\includegraphics[width=8cm]{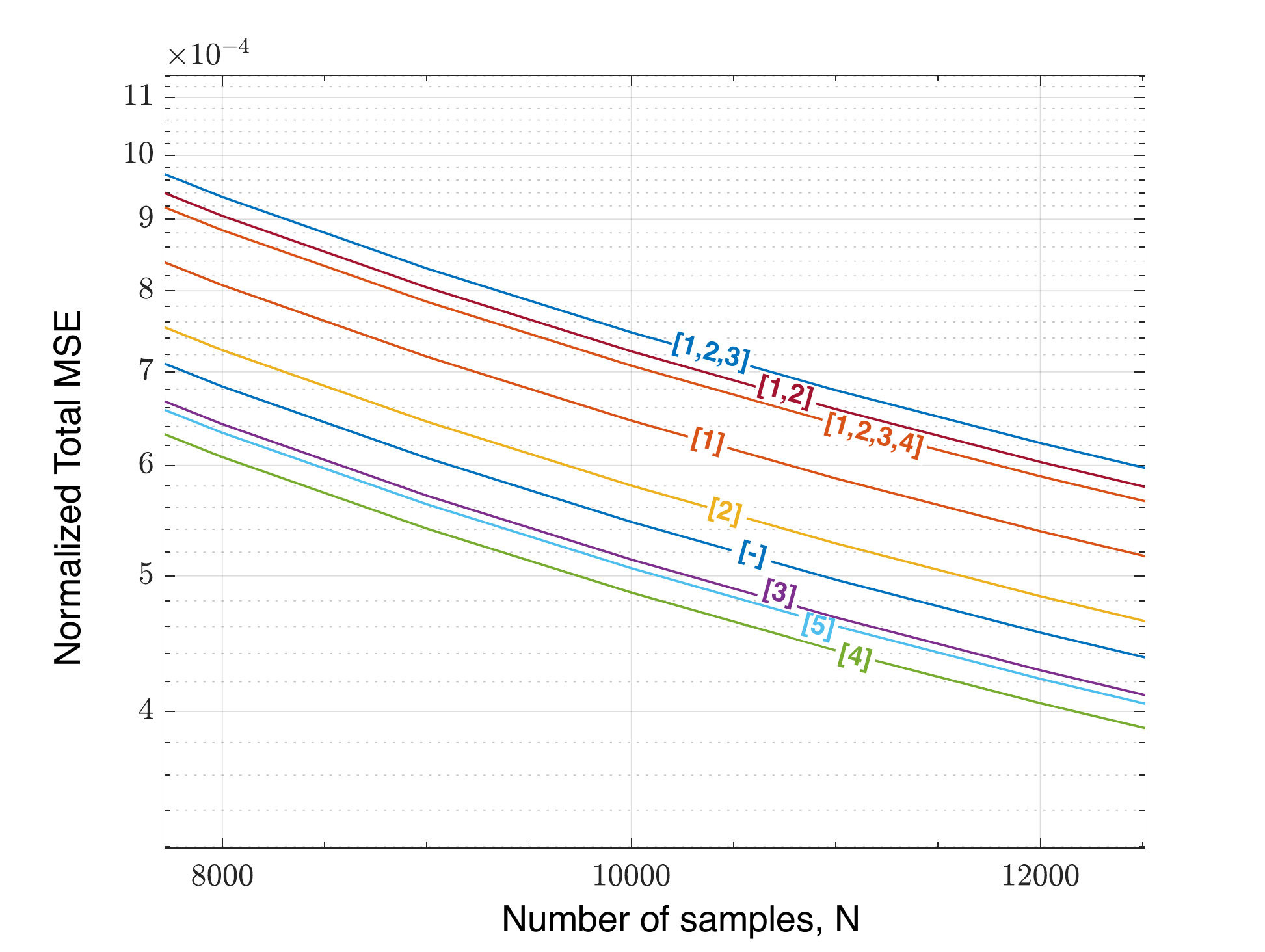}
		\label{fig:absence_b}
	}
	\caption{(a) Total MSE normalized by $c_{tot}^2$, as a function of varying number of samples $N$ for suboptimal unbiased estimator $\bm{\hat{c}}$ in the absence of different types of ligands indicated here by their indices. The region marked by the dashed rectangle is redrawn in (b) for better visualization.}
	\label{fig:absence}
\end{figure*}

\subsection{Effect of Absence of Ligands}
\label{sec:absence}
As a particular case, we investigate the estimation performance when the ligands of particular types, which were considered a priori in the implemented estimators, are not actually present in the channel. As discussed in Section \ref{sec:spectrumsensing}, the proposed estimators should be hardwired in the receiving cell with a set of ligand types potentially existing in the channel before its utilization in an application. Hence, hardwiring the estimator with a large set of ligand types might be necessary for an application medium that could potentially contain varying types of ligands. In these cases, it is likely that some of the considered ligand types are not present in the medium at the time of channel sensing. In order to analyze the effect of absence of any ligand type, we need to change the previously considered performance metric, i.e., average NMSE, because the concentration of these ligand types is effectively zero, and it is not plausible to normalize the MSE with a zero concentration. Instead, we define a new metric as the total MSE normalized by the square of total concentration, i.e., 
\begin{align}
	\MSE_{tot}[\bm{\hat{c}}]/c_{tot}^2 = \frac{1}{c_{tot}^2}\sum_{i=1}^{M} \MSE[\hat{c}_i].
\end{align}
This metric enables a fair assessment of the performance in the absence of ligands. 

For the analysis, we consider the default setting with $M=5$, $\chi = 5$, $k_M^- = 1\text{s}^{-1}$, $\nu = 3$ by leaving the number of samples $N$ as a variable. We investigate the cases when each one of the ligand types is absent, as well as the case when multiple ligands are absent at the same time. In all cases, the concentration ratios of the remaining ligand types are assumed to be equal. The results are shown in Fig. \ref{fig:absence_a} (with a magnified view provided in Fig. \ref{fig:absence_b}) for the unbiased estimator $\bm{\hat{c}}$, and compared to the default case when all types of ligands, that are initially hardwired to the estimator, are present. The numbers in square brackets indicate the index value $i$ of the ligand types varying in their unbinding rates, such that $k_i^- = \chi^{M-i} k_M = 5^{5-i}$ for $i \in \{1, \dots, 5\}$. As is seen, the estimation performance only slightly changes in different cases, and thus, the proposed estimators can be considered robust against the absence of any ligand types that are hardwired a priori. Although there is not a clear trend in the performance with varying types of ligands that are absent, we can see that while the absence of ligands with higher unbinding rates improve the overall estimation performance, the absence of ligands with higher affinity with receptors degrades the performance compared to the default case. 
\begin{figure*}[!t]
	\centering
	\subfigure[]{
		\includegraphics[width=5.5cm]{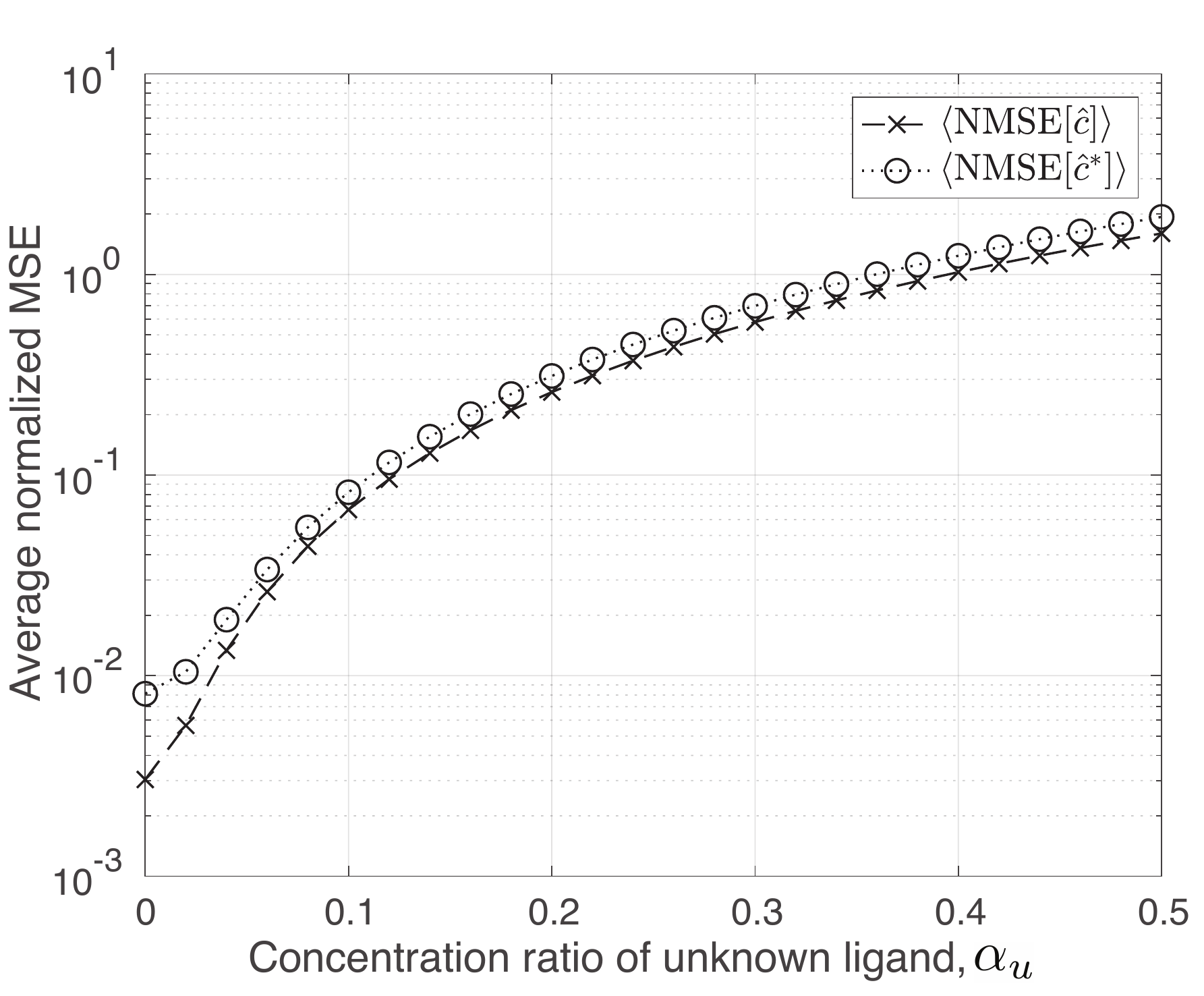}
		\label{fig:new_ligand_a}
	}
	\subfigure[]{
		\includegraphics[width=5.5cm]{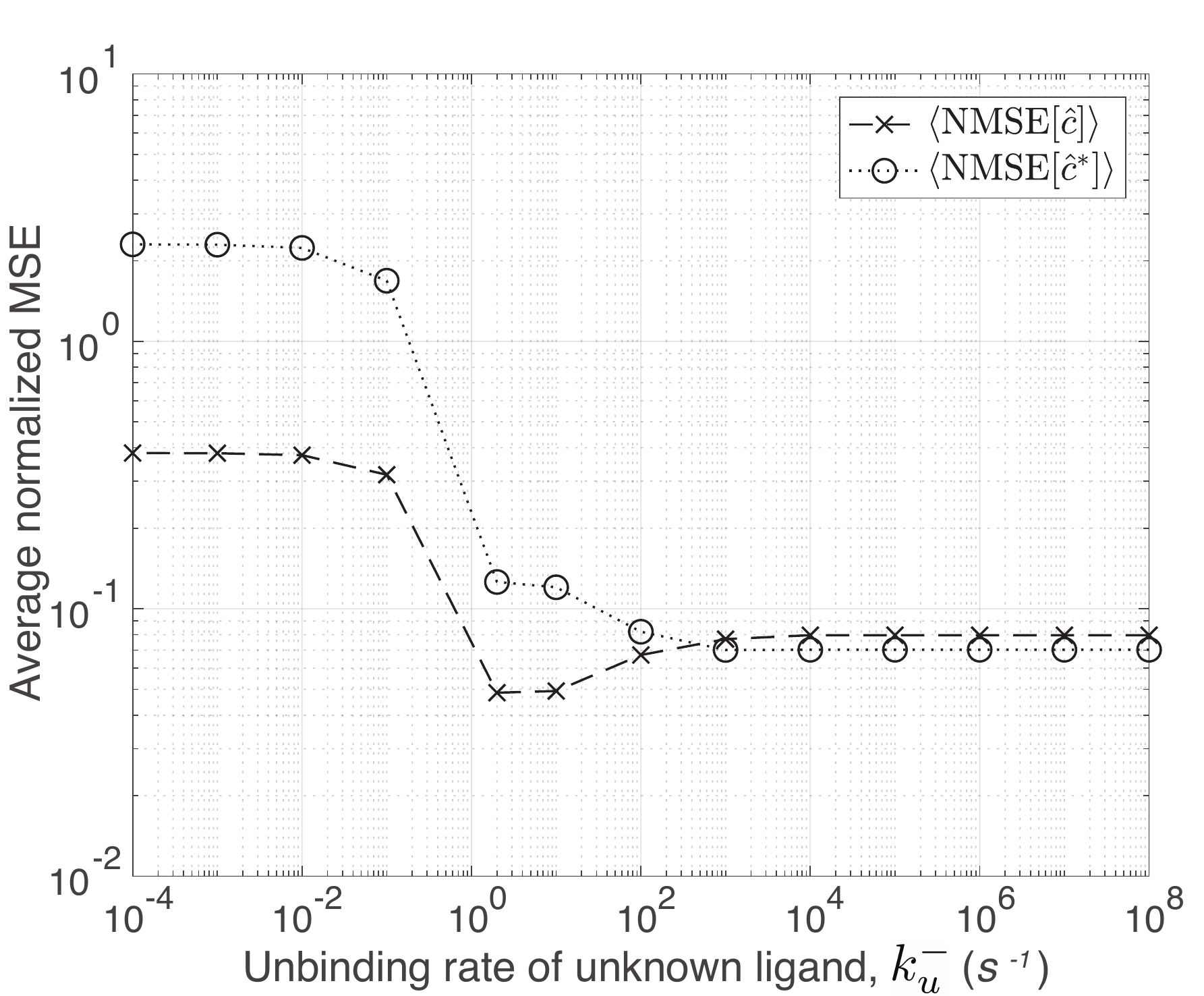}
		\label{fig:new_ligand_b}
	}
	\subfigure[]{
		\includegraphics[width=5.5cm]{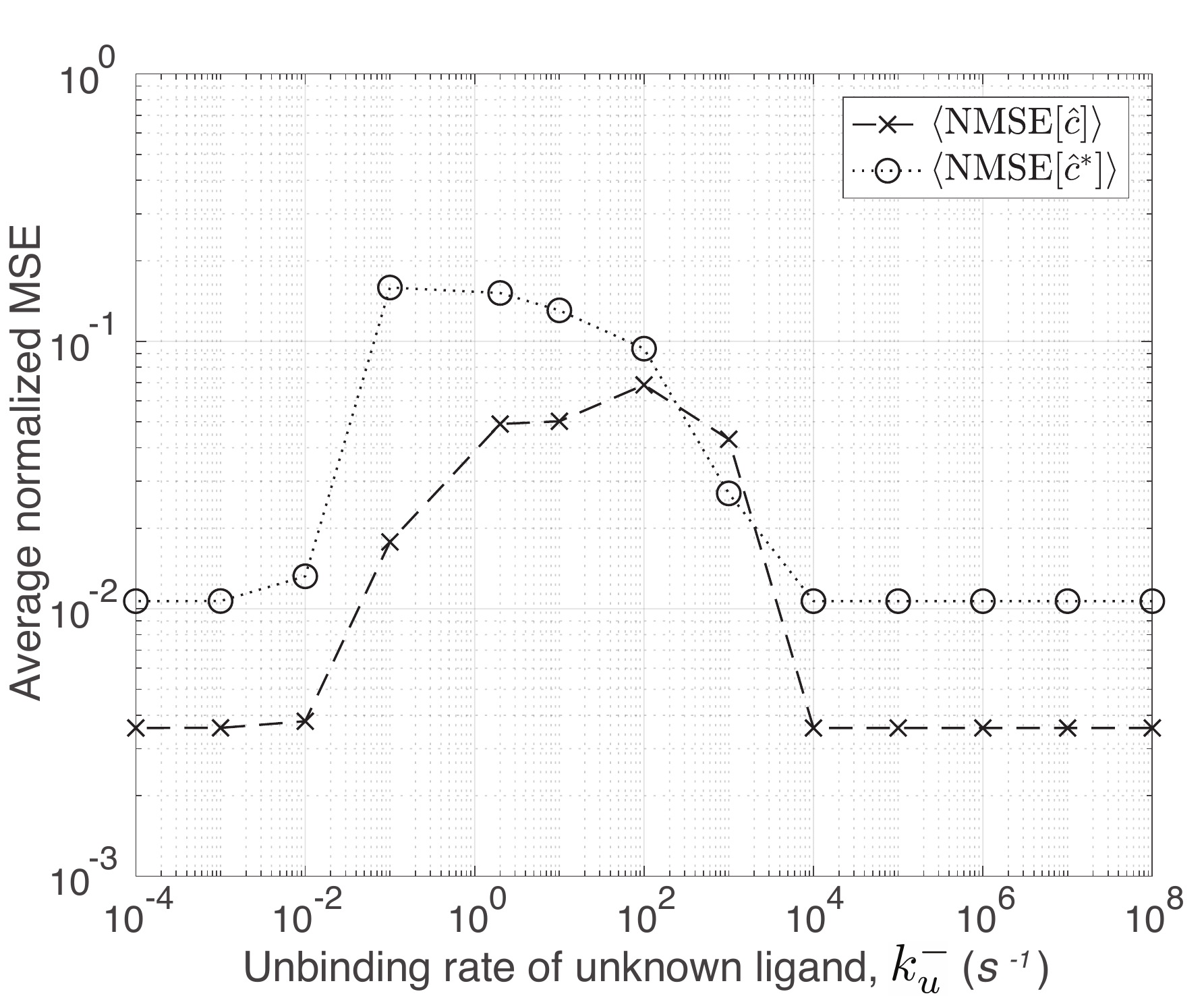}
		\label{fig:new_ligand_c}
	}
	\caption{Average NMSE in the presence of an unknown ligand  (a) with varying concentration ratio; (b) with varying unbinding rate; (c) with varying unbinding rate when very short and very long binding events are filtered out.}
	\label{fig:new_ligand}
\end{figure*}

\subsection{Effect of Unknown Ligand Types}
We also investigate the estimator performance when new types of ligands are introduced to the channel medium. Note that these new ligands are unknown to the estimators. This problem is relevant for communication media with a varying characteristics in terms of interferer molecule types. Our objective is to understand the effect of the unbinding rate and concentration ratio of the new ligand type on the performance of the estimators in estimating the concentration of the known ligands in terms of average NMSE. We derive the average NMSE in the case of new ligands in Appendix \ref{AppendixA}, and show that the unbiased estimator becomes biased in this case. The results of the analyses are provided in Fig. \ref{fig:new_ligand}. In addition to the unbiased estimator, which now becomes biased, we also provide results for simplified biased estimator introduced in Section \ref{sec:biased}. We use the default setting for existing ligands, i.e., $M = 5$, $k_M^- = 1\text{s}^{-1}$, $\chi = 5$, $\nu = 3$. Note that in this setting the unbinding rates of existing ligands become $k_5^- = 1\text{s}^{-1}$, $k_4^- = 5\text{s}^{-1}$, $k_3^- = 25\text{s}^{-1}$, $k_2^- = 125\text{s}^{-1}$, $k_1^- = 625\text{s}^{-1}$. The concentration of the existing ligands are considered to be equal, such that when a new type of ligand is introduced with a certain concentration ratio $\alpha_u$, the ratio of the existing ligand types becomes $\alpha_i = (1-\alpha_u)/M$ for $i \in \{1, \dots, M \}$. 

We first analyze the effect of the concentration ratio of the new ligand type, $\alpha_u$. We take its unbinding rate as $k_u^- = 100\text{s}^{-1}$, such that its affinity with the receptors is close to that of the existing ligands.  As is seen in Fig. \ref{fig:new_ligand_a}, the ratio of the introduced ligand has a substantial effect on the performance of both estimators, and the estimation becomes highly unreliable when $\alpha_u > 0.2$.  In Fig. \ref{fig:new_ligand_b}, we investigate the effect of unbinding rate $k_u^-$ of this new ligand while keeping its concentration ratio fixed at $\alpha_u = 0.1$. As is seen for both types of estimators, the effect of the unknown ligand on the performance is more pronounced when its unbinding rate is lower (i.e., its affinity is higher) than the existing ligands that are known to the estimators.

The detrimental effect of the unknown ligands can be reduced by adjusting the lower and upper time thresholds, i.e., $T_0$ and $T_M$ defined in \eqref{timethreshold}, to filter out the binding events that last substantially longer or shorter than those resulting from the known ligand types. For our analysis, we set $T_0 = T_1/5 = \nu/(5k_1^-) = [3/(5 \times 625\text{s}^{-1})]  = 960 \mu\text{s}$ and $T_M = T_{M-1}/5 = \nu/(5k_{M-1}^-) =  [3/(5 \times 5\text{s}^{-1})] = 120 \text{ms}$, such that the binding events that last shorter than $960 \mu\text{s}$ or longer than $120 \text{ms}$ are filtered out. We provide the results of our analysis in Fig. \ref{fig:new_ligand_c} for varying unbinding rate of the new ligand. As compared to the results in Fig. \ref{fig:new_ligand_b}, when the unbinding rate of the new ligand is significantly higher or lower than the unbinding rate of the existing ligands, its detrimental effect is removed. However, when the new ligand has  similar characteristics with the existing ligands, the effect cannot be removed. 

\section{Discussion on Implementation}
\label{sec:discussion}
\begin{figure*}[!t]
	\centering
	\includegraphics[width=17cm]{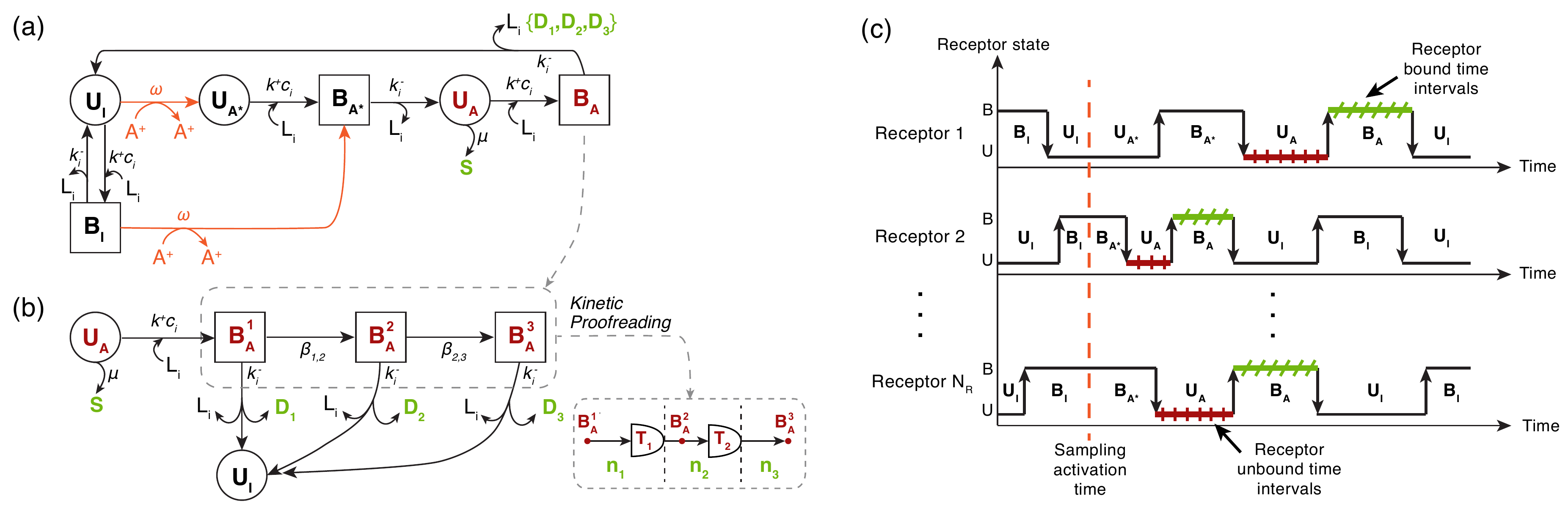}
	\caption{(a) State diagram of the proposed synthetic receptor design that transduces bound and unbound time durations of a receptor into second messengers for intracellular signal processing. The orange line indicates the state transitions initiated by the activation molecules. (b) A closer look into the KPR mechanism with the three KPR substates demonstrated with the corresponding state transition rates. (c) Demonstration of the activation and sampling cycles of the proposed receptors. Dashed orange line marks the reaction time of activation molecules with the receptors.}
	\label{fig:kpr}
\end{figure*}

In this section, we investigate the practical aspects of the proposed channel sensing methods. For the implementation of the method, we focus on a synthetic biology-based approach, because biosensor-based approaches for MC receiver, as overviewed in \cite{kuscu2018survey, kuscu2016physical}, do not allow inspecting the states of individual receptors, and thus, implementing the proposed estimators.

The key element in the channel sensing is the biological receptors, which are the interface between the exterior and interior of a living cell, and transduce the external signals represented by the concentration of ligands into intracellular signals in the form of concentration of second messengers inside a living cell. The transduced signals need to be further processed for the estimation to be achieved. 

The proposed estimators, both unbiased and biased, rely on two statistics, i.e., total unbound time $T_u$, and the number of binding events $n_i$ of durations within $[T_{i-1}, T_i]$ for $i \in \{1, \dots, M\}$. Our first aim is to provide a practical synthetic receptor design that can transduce both the unbound time and bound time information into the concentration of different intracellular molecules. We then investigate a chemical reaction network (CRN) that can chemically process these intracellular molecules to perform the calculations required for the proposed estimators. 

\subsection{Acquisition of Receptor Unbound/Bound Time Duration Statistics}
The proposed estimators require the sampling of only a single pair of unbound and bound time durations from each receptor, as demonstrated in Fig. \ref{fig:kpr}(c), because the information of the exact number of independent samples is crucial for the estimation performance. To equate the number of samples and receptors, i.e., $N = N_R$, we first propose a receptor activation mechanism that can be triggered by the receiver cell when it decides to sense the channel. In this scheme, the receptors can be in one of the 6 main states, i.e., inactive unbound/bound, active unbound/bound and intermediate unbound/bound states, depending on the history of their reactions with ligands and intracellular molecules. The receptors can perform the sampling of the unbound time durations only in the active unbound state, and the bound time durations only in the active bound state through different mechanisms, which will be discussed shortly. Next, we describe the proposed activation mechanism along with the sampling of unbound time durations, and then we propose a modified kinetic proofreading (KPR) scheme for the sampling of bound time durations. 

\subsubsection{\bf{Receptor Activation and Transduction of Total Unbound Time Duration}}
We propose a receptor activation mechanism to control the start time and duration of the channel sensing, such that only one unbound/bound time duration is sampled from each receptor. In this scheme, the sampling process starts with the generation of activation molecules $A^+$, produced by the cell in an impulsive manner, when the cell decides to sample the receptor states, as demonstrated in Fig. \ref{fig:kpr}. The generation of activation signal, thus, occurs in bursts, through the following reaction
\begin{equation}
	\ce{ {\varnothing}  ->[{s(t) \psi^+}] {A^+}},
\end{equation}
where the time-varying generation rate is given as $s(t) \psi^+$, with $s(t) \approx \delta(t-t_A)$ being a very short pulse signal centered around the activation time $t_A$. 
Shortly after activation, the cell generates deactivation molecules $A^-$, through the following reaction
\begin{equation}
	\ce{ {\varnothing}  ->[{d(t) \psi^-}] {A^-}}.
\end{equation}
The reaction rate is given by $d(t) \psi^-$, with $d(t) \approx \delta(t-t_D)$ being again an impulse-like signal centered around the deactivation time $t_D$. The generated deactivation molecules degrade the existing activation molecules within the cell at a rate $\rho$, i.e.,
\begin{equation}
\ce{ {A^+ + A^-} ->[{\rho}] {\varnothing} },
\end{equation}
such that the duration of the overall sampling process can be controlled.
The inactive receptors, i.e., $U_I$ and $B_I$, transition into their intermediate states,  i.e., $U_A^\ast$ and $B_A^\ast$, upon reacting with an activation molecule $A^+$ at a rate $\omega$, i.e.,
\begin{align}
	\ce{ {U_I + A^+} &->[{\omega}] {U_A^\ast + A^+} }\\ \nonumber
	\ce{ {B_I + A^+} &->[{\omega}] {B_A^\ast + A^+} }.
\end{align}
The binding of an unbound receptor in the intermediate state $U_A^\ast$, transforms it into an intermediate bound receptor $B_A^\ast$, i.e.,
\begin{equation}
	\ce{ {U_A^\ast + L_i} ->[{k^+c_i }] {B_A^\ast} }, 
\end{equation}
where $L_i$ denotes a ligand molecule of $i^\text{th}$ type. Upon the first unbinding event, a bound receptor in the intermediate state $B_A^\ast$ goes into the active unbound state $U_A$, i.e., 
\begin{equation}
	\ce{ {B_A^\ast} ->[{k_i^-}] {U_A + L_i} }.
\end{equation}
In the active unbound state $U_A$, the receptor produces the secondary messenger molecules $S$ at a constant rate through the following first-order reaction,
\begin{equation} \label{eq:generationofS}
\ce{ U_A   ->[{\mu}] {U_A + S}}.
\end{equation}
As a result of this reaction, the steady-state concentration of the produced $S$ molecules becomes proportional to the total unbound time $T_u$, as we will see in Section \ref{ss_analysis}.

Upon binding a ligand, the active unbound receptor $U_A$ switches into the first KPR substate of the active bound state $B_A^1$, i.e., 
\begin{equation}
	\ce{ {U_A + L_i}   ->[{k^+c_i}] {B_A^1}}.
\end{equation}
As a result, the modified KPR scheme, consisting of $M$ substates, $\{B_A^1, \dots, B_A^M\}$, becomes activated. 

We provide some examples in Fig. \ref{fig:kpr}(c) for receptor state trajectories governed by the proposed activation mechanism. Receptor 1 is in inactive unbound state when the activation signal is sent. The reaction with activation molecules $A^+$ turns it into intermediate unbound state $U_A^\ast$. Next, with the binding of a ligand, it goes into intermediate bound state. Following the unbinding of the bound ligand, it finally gets activated in the unbound state. During the active unbound state, it generates $S$ molecules following the reaction \eqref{eq:generationofS}. When it binds a ligand again, it switches into active bound state, where the KPR mechanism is activated. The next unbinding event brings the receptor back into the inactive unbound state. As such one cycle of sampling of unbound and bound time duration is completed. On the other hand, Receptor 2 is in the inactive bound state $U_I$ at the time of activation. The activation reaction switches it into the intermediate bound state $B_A^\ast$, during which it is still not able to generate any second messenger. After the first unbinding event it transitions into the active bound state $U_A$, where it generates $S$ molecules. Upon the next binding, it becomes active in the bound state $B_A$, and activates the KPR mechanism. This sampling cycle is also completed with the following unbinding event that leads it into the inactive unbound state $U_I$.  

To ensure that the inactivated receptors are not re-activated during the same sampling process for the sake of obtaining only a single pair of unbound and bound duration samples from each receptor, the generation rates of activation and deactivation molecules, i.e., $\psi^+$ and $\psi^-$, respectively, as well as the rate of reaction between activation molecules and receptors, i.e., $\omega$, and the rate of deactivation reaction $\rho$ should be very high compared to the ligand-receptor binding/unbinding reaction rates.

\subsubsection{\bf{Kinetic Proofreading and Transduction of Bound Time Durations}}
\label{kpr_scheme}
For the sampling of the bound time durations, we propose a modified KPR scheme. In the KPR mechanism, the active bound receptor sequentially visits its $M$ substates in an irreversible manner during the bound time period by undergoing a series of conformational changes at specific transition rates, as shown in Fig. \ref{fig:kpr}(b). In each internal state, the receptor can directly return to the initial inactive unbound state $U_I$ if the bound ligand unbinds from the receptor. In our modified KPR scheme, while returning to the initial unbound state, the receptor releases an intracellular molecule $D$, type of which is specific to the last occupied KPR substate. As the unbinding rate is different for each ligand type, the last occupied substate is informative of the type of the bound ligand in a probabilistic manner. This information is encoded into the number of $D_i$ molecules generated by all active bound receptors, which becomes proportional to the number of last visits made to the $B_A^i$ substate at steady-state, as discussed in Section \ref{ss_analysis}. 

In order for the proposed KPR scheme to provide the required statistics for the estimation of ligand concentration ratios, the transition rates, $\beta$'s, between substates should be set in accordance with the time thresholds introduced in \eqref{timethreshold}. As such, the resulting number of second messengers, $D_i$, produced from the internal states $B_A^i$ will approximate the actual number of binding events $n_i$ of durations within corresponding time ranges. The transition rates between the KPR states can be set as a function of time thresholds $T_i$'s as follows 
\begin{align}
\beta_{i,i+1} = \kappa_i/\left(T_i^- - T_{i-1}^-\right) ~~~ \text{for}~ i \in \{1, \dots,	 M-1\} ,
\end{align}
where $\kappa_i$'s are tuning parameters to adjust the transition rates. In the next section, we will show that setting $\kappa_i = 3/5$ provides a good approximation for the number of binding events falling in each time interval for the unbiased estimator, where $T_i = 3/k_i^-$ for $i \in \{1, \dots,	 M-1\}$, and $T_0 = 0$. 

\subsubsection{\bf{Steady-State Analysis}}
\label{ss_analysis}
\begin{figure}[!b]
	\centering
	\includegraphics[width=7cm]{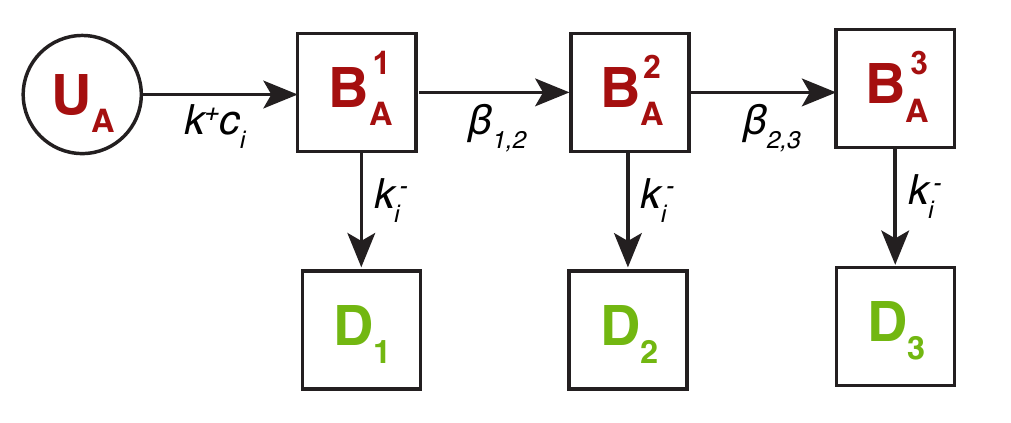}
	\caption{The kinetic scheme of an active receptor given as a Markov Process. }
	\label{fig:kinetic}
\end{figure}

We can now provide a steady-state analysis for the transduction of unbound and bound time durations of receptors into second messengers, i.e., $S$ and $D$ molecules. We consider the case when there are $M=3$ different types of ligands co-existing in the channel, such that each of the receptors has three KPR substates, as shown in Fig. \ref{fig:kpr}(a). For the sake of brevity of the analysis, we omit the activation mechanism, and focus only on the active receptors. The considered system for steady-state analysis is then a kinetic scheme of Markovian nature, and redrawn in Fig. \ref{fig:kinetic}, demonstrating possible states of active receptors along with the relevant transition rates.

In order to write the chemical master equation (CME) for this kinetic scheme, at the moment, we consider the case of single type of ligands. The CME can then be given as a set of differential equations, i.e., 
\begin{align} \label{eq:KPRode}
\frac{dP_{U_A|i}}{dt} &= - k^+ c_i P_{U_A|i} \\ \nonumber
\frac{dP_{B_A^1|i}}{dt} &= k^+ c_i P_{U_A|i}  - \beta_{1,2} P_{B_A^1|i} - k_i^- P_{B_A^1|i} \\ \nonumber
\frac{dP_{B_A^2|i}}{dt} &=  \beta_{1,2} P_{B_A^1|i} - \beta_{2,3} P_{B_A^2|i}- k_i^- P_{B_A^2|i} \\ \nonumber
\frac{dP_{B_A^3|i}}{dt} &=  \beta_{2,3} P_{B_A^2|i} - k_i^- P_{B_A^3|i}  \\ \nonumber
\frac{dP_{D_j|i}}{dt} &=  k_i^- P_{B_A^j|i}~~~ \text{for}~ j \in \{1, \dots,	 M \},
\end{align}
where $P_{U_A|i}, P_{B_A^1|i}, P_{B_A^2|i}, P_{B_A^3|i}$ are the time-varying probabilities of an active receptor to be in the unbound state, and in each of the KPR substates, respectively, conditioned on the presence of only the $i^\text{th}$ type of ligand. $D_j$'s represent virtual states of absorbing nature for an active receptor, and $P_{D_j|i}$ denotes the probability of an active receptor to generate an intracellular $D_j$ molecule and return to the inactive unbound state when an $i^\text{th}$ type of ligand is bound. The escape rate from KPR substates is equal to the unbinding rate of the bound ligand, $k_i^-$.
\begin{figure*}[!t]
	\centering
	\includegraphics[width=18cm]{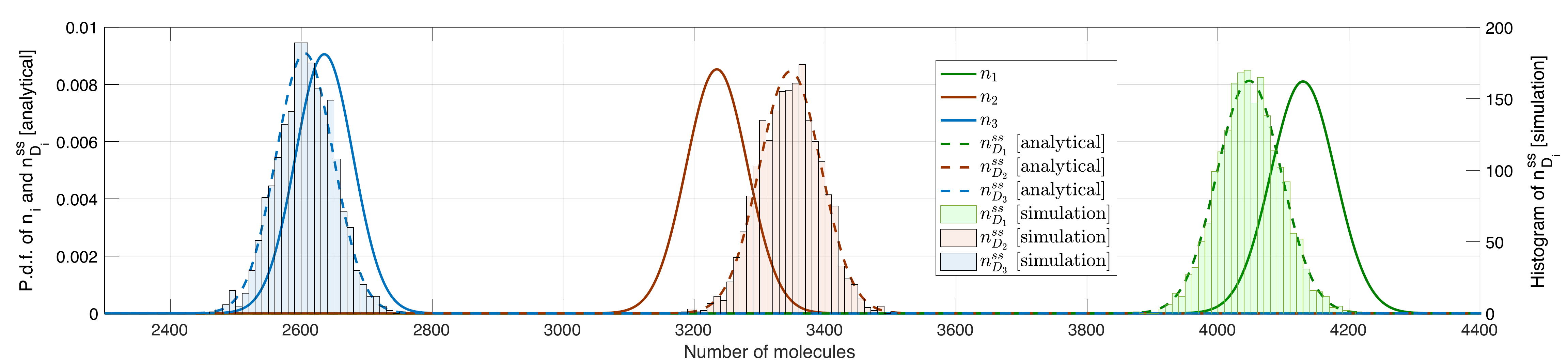}
	\caption{\textbf{Solid lines:} Gaussian-approximated probability distribution of number of binding events $n_i$ that fall in each time interval defined by the time thresholds $T_i$'s given according to \eqref{timethreshold} with $\nu = 3$. \textbf{Dashed lines:} Steady-state probability distribution of number of $D$ molecules $n_{D_j}^{ss}$ for $\kappa = 3/5$, given in \eqref{eq:Ddistribution}. \textbf{Histogram:} Steady-state probability distribution of number of $D$ molecules $n_{D_j}^{ss}$ for $\kappa = 3/5$ obtained through a Monte Carlo simulation of the kinetic scheme demonstrated in Fig. \ref{fig:kinetic}.
	}
	\label{fig:KPR_analysis}
\end{figure*}

The steady-state solution of the CME in \eqref{eq:KPRode} is analytically obtained with the initial conditions  $P_{U_A|i}^0 = 1$, $P_{B_A^1|i}^0 = P_{B_A^2|i}^0 = P_{B_A^3|i}^0  = P_{D_1|i}^0 = P_{D_2|i}^0= P_{D_3|i}^0 = 0$, as follows
\begin{align}
P_{U_A|i}^{ss} &= P_{B_A^1|i}^{ss} = P_{B_A^2|i}^{ss} = P_{B_A^3|i}^{ss} =  0, \\ \nonumber
P_{D_1|i}^{ss} &=  \frac{k_i^-}{\beta_{1,2} + k_i^-}, \\ \nonumber
P_{D_2|i}^{ss} &=  \frac{\beta_{1,2} k_i^-}{\beta_{1,2} \beta_{2,3} + \beta_{1,2} k_i^- + \beta_{2,3} k_i^- + (k_i^-)^2}, \\ \nonumber
P_{D_3|i}^{ss} &=  \frac{\beta_{1,2} \beta_{2,3}}{\beta_{1,2} \beta_{2,3} + \beta_{1,2} k_i^- + \beta_{2,3} k_i^- + (k_i^-)^2}.
\end{align}
In the presence of three types of ligands, the overall steady-state probabilities can be written as follows
\begin{align}
	P_{U_A}^{ss} &= P_{B_A^1}^{ss} = P_{B_A^2}^{ss} = P_{B_A^3}^{ss} =  0, \\ \nonumber
	P_{D_j}^{ss} &= \sum_{i=1}^{M=3} \alpha_i P_{D_j|i}^{ss}.
\end{align}

Given that all active receptors independently follow the same kinetic scheme, and assuming that each receptor goes through the active state for once during a sampling process, the mean number of generated intracellular $D$ molecules at steady-state can be given as
\begin{equation} \label{D_ss}
\E[n_{D_j}^{ss}] = N P_{D_j}^{ss},~~ \text{for}~ j \in \{1, \dots,	 M=3 \}.
\end{equation}
Here we use our previous assumption that number of samples is equal to the number of receptors, i.e., $N = N_R$. Given the statistical independence of receptors, we can also write the variance of number of $D$ molecules as follows
\begin{equation}
\Var[n_{D_j}^{ss}] = N P_{D_j}^{ss} (1-P_{D_j}^{ss}),~~ \text{for}~ j \in \{1, \dots,	 M=3 \}.
\end{equation}
Assuming that the number of receptors is sufficiently high ($N_R = N = 10000$ in the considered case), we can approximate the random number of produced $D$ molecules at steady-state with a Gaussian distribution, i.e., 
\begin{equation}
n_{D_j}^{ss} \sim \mathcal{N}\left( \E[ n_{D_j}^{ss}], \Var[n_{D_j}^{ss}] \right),~~ \text{for}~ j \in \{1, \dots,	 M=3 \}.
\label{eq:Ddistribution}
\end{equation}

As discussed in Section \ref{kpr_scheme}, the transition rates, $\beta$'s, between KPR substates should be optimized for obtaining the most accurate representation of actual number of binding events $n_i$'s with $D$ molecules. However, for the sake of brevity of this discussion, we leave the optimization problem as a future research. Here we provide the steady-state probability distribution of number of $D$ molecules $n_{D_j}^{ss}$ for $\kappa = 3/5$ in Fig. \ref{fig:KPR_analysis}. In the same figure, the results are compared to the histogram of the same statistic obtained through a Monte Carlo simulation of the kinetic scheme demonstrated in Fig. \ref{fig:kinetic}, and the probability distribution of number of binding events $n_i$ that fall in each time interval defined by the time thresholds $T_i$'s given according to \eqref{timethreshold} with $\nu = 3$. Here, assuming that $N$ is large enough, we also approximate the binomial distribution of $n_i$ with Gaussian distribution, i.e., $n_i \sim \mathcal{N}\left(\E[n_i], \Var[n_i] \right)$, where the mean and variance of $n_i$ are given in \eqref{eq:meann} and \eqref{eq:varn}, respectively. As is seen, the analytical results are in very good match with the simulation results. They also show that the KPR scheme with the selected transition rates can approximate the number of binding events in each time interval. However, the results also imply that the transition rates should be further optimized to obtain better approximation.

The generation of intracellular $S$ molecules encoding the total unbound time duration, on the other hand, is governed by the following rate equations, 
\begin{align} \label{rate_equation}
\frac{d\E[n_{U_A}]}{dt} &= N \frac{dP_{U_A}}{dt} = - k^+ c_{tot} \E[n_{U_A}], \\ \nonumber
\frac{d\E[n_S]}{dt} &= - \mu \E[n_{U_A}].
\end{align}
In writing \eqref{rate_equation}, for mathematical convenience, we consider the receptors as if they independently start in the active unbound state at the same time. This assumption does not degrade the accuracy of the analysis, because the generated $S$ molecules, whose generation rate is dependent on the number of active unbound receptors, are not degraded throughout the entire process (same as $D$ molecules), and we are only concerned about the steady-state statistics of the intracellular molecules, and not interested in their time-varying statistics. Hence, the steady-state solution of \eqref{rate_equation} for the initial conditions $\E[n_{U_A}^0] = N$ and $\E[n_S^0]  = 0$, is given as
\begin{align} \label{U_ss}
\E[n_{U_A}^{ss}] = 0.
\end{align}
\begin{align} \label{S_ss}
\E[n_{S}^{ss}] =  \frac{\mu N }{k^+ c_{tot}}.
\end{align}

The produced $D$ and $S$ molecules, whose expected numbers at steady-state are given in \eqref{D_ss} and \eqref{S_ss}, respectively, will be input to the estimator CRN introduced in the next section.

\subsection{Estimation with Chemical Reaction Networks}
Once the transduction of total unbound time $T_u$ and the number of binding events $n_i$ is completed, the arithmetic operations required for the estimator can be realized through intracellular CRNs that can perform analog computations \cite{daniel2013synthetic}. Here we focus on the unbiased estimator; thus, the objective is to implement the following equation with a CRN: 
\begin{align} \label{c_estimate}
\hat{c_l} &= \frac{N-1}{N} \frac{1}{k^+ T_u}  \sum_{i=1}^M n_i w_{l,i}, \\ \nonumber
&\approx \frac{1}{k^+ T_u}  \sum_{i=1}^M n_i w_{l,i}, ~~~ \text{for}~ N \gg 1.
\end{align}
Accordingly, we need to obtain the weighted sum of number of $M$ different types of second messengers $D_i$'s corresponding to the number of binding events that fall in each time interval, divided by the concentration of $S$ molecules encoding the total unbound time duration. This can be achieved through the following CRN, which is designed based on the methodology introduced in \cite{chou2017chemical}: 
\begin{equation}
\ce{ {D_i} ->[{w_{j,i}}] {D_i} + { Y_j} },~~~ \text{for}~ i,j \in \{1, \dots,	 M=3 \},
\end{equation}
\begin{equation}
\ce{ S + Y_j ->[{k^+}] S },~~~ \text{for}~ j \in \{1, \dots,	 M=3 \}.
\end{equation}
In this CRN, while $Y$ molecules are generated by $D$ molecules with different rates set by the matrix $\bm{W} = \bm{S}^{-1}$ (see \eqref{Wmatrix}), they are consumed by the $S$ molecules that encode the total unbound time duration. The rate equation of the above CRN for the $i^\text{th}$ ligand can be written as
\begin{align}
\frac{d\E[ n_{Y_i}] }{dt} = \sum_{j=1}^M w_{i,j} \E[n_{D_j}]  &- k^+ \E[n_{S}]  \E[ n_{Y_i}], \\ \nonumber
&~~~ \text{for}~ i \in \{1, \dots,	 M = 3 \}.
\end{align}
Given the initial condition  $\E[n_{Y_i}^0] = 0$, the steady-state solution for $\E[n_{Y_i}]$ is obtained as
\begin{equation} \label{Y_ss}
\E[n_{Y_i}^{ss}] = \frac{1}{k^+ \E[n_{S}^{ss}] } \sum_{j=1}^M w_{i,j} \E[n_{D_j}^{ss}].  
\end{equation}
Recall from \eqref{Tu} that $\E[1/T_u] = (k^+ c_{tot})/(N-1) \approx (k^+ c_{tot})/N$ for large $N$. By combining this with \eqref{S_ss}, we can see that $1/(k^+ \E[n_{S}^{ss}]) = c_{tot}/(\mu N) \approx \E[1/( \mu k^+ T_u)]$, is on average proportional to the first part of \eqref{c_estimate}, i.e., $1/(k^+ T_u)$. The proportionality constant is the rate $\mu$, which is the generation rate of $S$ molecules and can be simply set to $1\text{s}^{-1}$ for a better approximation. Given that the steady-state number of $D_i$ molecules $n_{D_i}^{ss}$ approximates the actual number of binding events $n_i$, the mean number of $Y_i$ molecules at steady-state given in \eqref{Y_ss} becomes proportional to the concentration estimate of $i^\text{th}$ type of ligand $\hat{c}_i$ given in \eqref{c_estimate}.

We note that once the estimation through CRN is completed, the produced intracellular molecules, i.e., $D$, $S$ and $Y$ molecules, should be removed from the cell through a chemical degradation reaction before the cell performs the next round of channel sensing. The rate of the degradation reaction can be set according to the required frequency of channel sensing. 

\section{Conclusion}
\label{sec:conclusion}
In this paper, we develop channel sensing techniques for MC with ligand receptors for the first time in the literature. In light of the results, we discuss that the proposed technique can be utilized for developing reliable MC detection, modulation, and multiple access methods, as it proved effective in sensing the individual concentrations of multiple ligand types by using only a single type of receptors. The technique is practical and low-complexity, and can be implemented in resource-limited synthetic biological MC devices, e.g., engineered bacteria. In this direction, we also discuss a synthetic receptor design, built upon the kinetic proofreading mechanism, that can transduce the required statistics of ligand-receptor binding reaction into intracellular signals. Lastly, we discuss a chemical reaction network that can perform the required arithmetic operations. This study is not exhaustive, and there remain many challenges and opportunities calling for future research. For example, interesting generalizations can be made by studying non-equilibrium cases where the concentration of ligands are changing more rapidly compared to the binding kinetics. As discussed throughout the paper, the study can be extended with the applications of the proposed method in developing reliable detection methods for CSK, MoSK and RSK modulated MC signals, and molecular division multiple access techniques. 

\appendices
\section{Introducing Unknown Ligand Types to the Channel}
\label{AppendixA}
In the proposed suboptimal estimators, the time thresholds, $T_i$'s, and the corresponding $S$ and $H$ matrices, given in \eqref{eq:Smatrix} and \eqref{eq:Hmatrix}, respectively, are constructed assuming that there are $M$ types of ligands with the unbinding rates known to the receiver. Here, we investigate the case when $L$ different types of additional ligands with unbinding rates unknown to the receiver are introduced to the channel. We will derive the MSE for the suboptimal unbiased concentration estimator introduced in Section \ref{sec:unbiased}. The derivation of the biased estimator, investigated in Section \ref{sec:biased}, can be done in a similar way. We will see that in the case of unknown ligands, the unbiased estimator becomes biased. 

Since the receiver assumes that there are $M$ different types of ligands, the time domain is divided into $M$ different regions. When there are $L$ additional ligand types, the probability of a binding duration to fall in a specific time interval can be rewritten in vector form as follows
\begin{equation}
\bm{\p} =  \frac{\bm{\E[n]}}{N} =  \bm{S_r} \bm{\alpha_r}, \label{eq:appendixP}
\end{equation}
where $\bm{\p}$ is $(M \times 1)$ probability vector, and $\bm{\alpha_r}$ is an ($[M+L] \times 1$) vector of concentration ratios of all ligand types including the additional ones. Note that $\bm{\alpha_r}^T = [\bm{\alpha}^T  \alpha_{M+1}  \dots \alpha_{M+L} ]$, with $\bm{\alpha}$ is the $(M \times 1)$ vector of concentration ratios of known ligand types.  Here, $\bm{S_r}$ is an ($M \times [M+L]$) matrix, whose elements are given by
\begin{align}
\bm{S_r}(i,j) &= \e^{-(k^-_j T_{i-1})} - \e^{-(k^-_j T_{i})} \\ \nonumber
&~~~ \text{for}~ i \in \{1, \dots,	 M \}, ~ j \in \{1, \dots,	 M+L \}.
\end{align}

Using the knowledge of only $M$ ligand types, the receiver utilizes the concentration ratio estimator given in \eqref{Wmatrix}, as follows
\begin{align} \label{WmatrixApp}
\bm{\hat{\alpha}} = \left(\frac{1}{N}\right) \bm{W} \bm{n},  
\end{align}
where $\bm{W} = \bm{S}^{-1}$ is the inverse of $\bm{S}$ matrix, that is given in \eqref{eq:Smatrix}. The mean of the ratio estimator then becomes
\begin{equation}
\bm{\E[\hat{\alpha}]}=  \left( \frac{1}{N}\right) \bm{W} \bm{\E[n]} = \bm{W} \bm{\p} = \bm{S^{-1}} \bm{S_r} \bm{\alpha_r}. \label{eq:appendixMean}
\end{equation}
The bias of the ratio estimator can then be written as: 
\begin{align}
\bm{\Delta[\hat{\alpha}]} &= \bm{\E[\hat{\alpha}]} - \bm{\alpha} \\ \nonumber
&= \bm{S^{-1}} \bm{S_r} \bm{\alpha_r} - \bm{\alpha}.
\end{align}

The concentration estimator for individual ligand types is given as $\bm{\hat{c}} = \bm{\hat{\alpha}} \hat{c}_{tot}$. Note that the ML estimator of total ligand concentration, $\hat{c}_{tot}$, is unbiased. Therefore, the bias of the concentration estimator can be computed as follows
\begin{align}
\bm{\Delta[\hat{c}]} =  \bm{\Delta[\hat{\alpha}]} c_{tot},
\end{align}
where the total ligand concentration is now given as $c_{tot} = \sum_{i=1}^{M+L} c_i$. Recall from \eqref{eq:totalvarianceunbiased} that the variance of the concentration estimator is written as
\begin{align} \label{eq:totalvarianceunbiased2}
\bm{\Var[\hat{c}]} &= \Var[\hat{c}_{tot}] \bm{\Var[\hat{\alpha}]} + \Var[\hat{c}_{tot}] \left( \bm{\E[\hat{\alpha}]} \odot \bm{\E[\hat{\alpha}]}\right) \\ \nonumber
&+ \bm{\Var[\hat{\alpha}]} \E[\hat{c}_{tot}]^2.
\end{align}
Here, $\Var[\hat{c}_{tot}] = \frac{c^2_{tot}}{N-2}~~ \text{for} ~~ N > 2$ and $\E[\hat{c}_{tot}] = c_{tot}$. The variance of the ratio estimator, $\bm{\Var[\bm{\hat{\alpha}}]}$, can be calculated by using \eqref{eq:est_variance} and \eqref{eq:est_covariance}, with the new probability vector $\bm{\p}$, given in \eqref{eq:appendixP}.

Finally, the MSE of the concentration estimator in case of additional unknown ligands can be written as
\begin{equation}
\bm{\MSE[\hat{c}]}=  \bm{\Var[\hat{c}]} + \left( \bm{\Delta[\hat{c}]} \odot \bm{\Delta[\hat{c}]} \right).
\end{equation}

\bibliographystyle{ieeetran}
\bibliography{references}

\begin{thebibliography}{10}
\providecommand{\url}[1]{#1}
\csname url@samestyle\endcsname
\providecommand{\newblock}{\relax}
\providecommand{\bibinfo}[2]{#2}
\providecommand{\BIBentrySTDinterwordspacing}{\spaceskip=0pt\relax}
\providecommand{\BIBentryALTinterwordstretchfactor}{4}
\providecommand{\BIBentryALTinterwordspacing}{\spaceskip=\fontdimen2\font plus
\BIBentryALTinterwordstretchfactor\fontdimen3\font minus
  \fontdimen4\font\relax}
\providecommand{\BIBforeignlanguage}[2]{{%
\expandafter\ifx\csname l@#1\endcsname\relax
\typeout{** WARNING: IEEEtran.bst: No hyphenation pattern has been}%
\typeout{** loaded for the language `#1'. Using the pattern for}%
\typeout{** the default language instead.}%
\else
\language=\csname l@#1\endcsname
\fi
#2}}
\providecommand{\BIBdecl}{\relax}
\BIBdecl

\bibitem{akan2017fundamentals}
O.~B. Akan, H.~Ramezani, T.~Khan, N.~A. Abbasi, and M.~Kuscu, ``Fundamentals of
  molecular information and communication science,'' \emph{Proceedings of the
  IEEE}, vol. 105, no.~2, pp. 306--318, 2017.

\bibitem{akyildiz2010internet}
I.~F. Akyildiz and J.~M. Jornet, ``The internet of nano-things,'' \emph{IEEE
  Wireless Communications}, vol.~17, no.~6, 2010.

\bibitem{akyildiz2015internet}
I.~F. Akyildiz, M.~Pierobon, S.~Balasubramaniam, and Y.~Koucheryavy, ``The
  internet of bio-nano things,'' \emph{IEEE Communications Magazine}, vol.~53,
  no.~3, pp. 32--40, 2015.

\bibitem{kuscu2016internet}
M.~Kuscu and O.~B. Akan, ``The internet of molecular things based on fret,''
  \emph{IEEE Internet of Things Journal}, vol.~3, no.~1, pp. 4--17, 2016.

\bibitem{kuscu2018survey}
M.~Kuscu, E.~Dinc, B.~A. Bilgin, H.~Ramezani, and O.~B. Akan, ``Transmitter and
  receiver architectures for molecular communications: A survey on physical
  design with modulation, coding and detection techniques,'' \emph{Proceedings
  of the IEEE}, In press.

\bibitem{nakano2008microplatform}
T.~Nakano, Y.-H. Hsu, W.~C. Tang, T.~Suda, D.~Lin, T.~Koujin, T.~Haraguchi, and
  Y.~Hiraoka, ``Microplatform for intercellular communication,'' in \emph{2008
  3rd IEEE International Conference on Nano/Micro Engineered and Molecular
  Systems}.\hskip 1em plus 0.5em minus 0.4em\relax IEEE, 2008, pp. 476--479.

\bibitem{krishnaswamy2013time}
B.~Krishnaswamy, C.~M. Austin, J.~P. Bardill, D.~Russakow, G.~L. Holst, B.~K.
  Hammer, C.~R. Forest, and R.~Sivakumar, ``Time-elapse communication:
  Bacterial communication on a microfluidic chip,'' \emph{IEEE Transactions on
  Communications}, vol.~61, no.~12, pp. 5139--5151, 2013.

\bibitem{grebenstein2019biological}
L.~Grebenstein, J.~Kirchner, R.~S. Peixoto, W.~Zimmermann, F.~Irnstorfer,
  W.~Wicke, A.~Ahmadzadeh, V.~Jamali, G.~Fischer, R.~Weigel \emph{et~al.},
  ``Biological optical-to-chemical signal conversion interface: A small-scale
  modulator for molecular communications,'' \emph{IEEE transactions on
  nanobioscience}, vol.~18, no.~1, pp. 31--42, 2019.

\bibitem{dinc2017theoretical}
E.~Dinc and O.~B. Akan, ``Theoretical limits on multiuser molecular
  communication in internet of nano-bio things,'' \emph{IEEE Transactions on
  Nanobioscience}, vol.~16, no.~4, pp. 266--270, 2017.

\bibitem{alizadeh2018coexistence}
A.~Alizadeh, H.~R. Bahrami, M.~Maleki, N.~H. Tran, and P.~Mohseni, ``On the
  coexistence of nano networks: Sensing techniques for molecular
  communications,'' \emph{IEEE Transactions on Molecular, Biological and
  Multi-Scale Communications}, 2018.

\bibitem{jamali2016channel}
V.~Jamali, A.~Ahmadzadeh, C.~Jardin, H.~Sticht, and R.~Schober, ``Channel
  estimation for diffusive molecular communications,'' \emph{IEEE Transactions
  on Communications}, vol.~64, no.~10, pp. 4238--4252, 2016.

\bibitem{noel2015joint}
A.~Noel, K.~C. Cheung, and R.~Schober, ``Joint channel parameter estimation via
  diffusive molecular communication,'' \emph{IEEE Transactions on Molecular,
  Biological and Multi-Scale Communications}, vol.~1, no.~1, pp. 4--17, 2015.

\bibitem{bialek2012biophysics}
W.~Bialek, \emph{Biophysics: Searching for Principles}.\hskip 1em plus 0.5em
  minus 0.4em\relax Princeton University Press, 2012.

\bibitem{pan2013molecular}
A.~C. Pan, D.~W. Borhani, R.~O. Dror, and D.~E. Shaw, ``Molecular determinants
  of drug--receptor binding kinetics,'' \emph{Drug discovery today}, vol.~18,
  no. 13-14, pp. 667--673, 2013.

\bibitem{mora2015physical}
T.~Mora, ``Physical limit to concentration sensing amid spurious ligands,''
  \emph{Physical Review Letters}, vol. 115, no.~3, p. 038102, 2015.

\bibitem{deng2015modeling}
Y.~Deng, A.~Noel, M.~Elkashlan, A.~Nallanathan, and K.~C. Cheung, ``Modeling
  and simulation of molecular communication systems with a reversible
  adsorption receiver,'' \emph{IEEE Transactions on Molecular, Biological and
  Multi-Scale Communications}, vol.~1, no.~4, pp. 347--362, 2015.

\bibitem{ahmadzadeh2016comprehensive}
A.~Ahmadzadeh, H.~Arjmandi, A.~Burkovski, and R.~Schober, ``Comprehensive
  reactive receiver modeling for diffusive molecular communication systems:
  Reversible binding, molecule degradation, and finite number of receptors,''
  \emph{IEEE Transactions on Nanobioscience}, vol.~15, no.~7, pp. 713--727,
  2016.

\bibitem{kuscu2016modeling}
M.~Kuscu and O.~B. Akan, ``Modeling and analysis of sinw fet-based molecular
  communication receiver,'' \emph{IEEE Transactions on Communications},
  vol.~64, no.~9, pp. 3708--3721, 2016.

\bibitem{chou2015markovian}
C.~T. Chou, ``A markovian approach to the optimal demodulation of
  diffusion-based molecular communication networks,'' \emph{IEEE Transactions
  on Communications}, vol.~63, no.~10, pp. 3728--3743, 2015.

\bibitem{endres2009maximum}
R.~G. Endres and N.~S. Wingreen, ``Maximum likelihood and the single
  receptor,'' \emph{Physical Review Letters}, vol. 103, no.~15, p. 158101,
  2009.

\bibitem{chou2015maximum}
C.~T. Chou, ``Maximum a-posteriori decoding for diffusion-based molecular
  communication using analog filters,'' \emph{IEEE Transactions on
  Nanotechnology}, vol.~14, no.~6, pp. 1054--1067, 2015.

\bibitem{kuscu2018maximum}
M.~Kuscu and O.~B. Akan, ``Maximum likelihood detection with ligand receptors
  for diffusion-based molecular communications in internet of bio-nano
  things,'' \emph{IEEE Transactions on Nanobioscience}, vol.~17, no.~1, pp.
  44--54, 2018.

\bibitem{singh2017simple}
V.~Singh and I.~Nemenman, ``Simple biochemical networks allow accurate sensing
  of multiple ligands with a single receptor,'' \emph{PLoS Computational
  Biology}, vol.~13, no.~4, p. e1005490, 2017.

\bibitem{lalanne2015chemodetection}
J.-B. Lalanne and P.~Fran{\c{c}}ois, ``Chemodetection in fluctuating
  environments: Receptor coupling, buffering, and antagonism,''
  \emph{Proceedings of the National Academy of Sciences}, vol. 112, no.~6, pp.
  1898--1903, 2015.

\bibitem{siggia2013decisions}
E.~D. Siggia and M.~Vergassola, ``Decisions on the fly in cellular sensory
  systems,'' \emph{Proceedings of the National Academy of Sciences}, vol. 110,
  no.~39, pp. E3704--E3712, 2013.

\bibitem{mckeithan1995kinetic}
T.~W. Mckeithan, ``Kinetic proofreading in t-cell receptor signal
  transduction,'' \emph{Proceedings of the National Academy of Sciences},
  vol.~92, no.~11, pp. 5042--5046, 1995.

\bibitem{muzio2018selective}
G.~Muzio, M.~Kuscu, and O.~B. Akan, ``Selective signal detection with ligand
  receptors under interference in molecular communications,'' in \emph{2018
  IEEE 19th International Workshop on Signal Processing Advances in Wireless
  Communications (SPAWC)}.\hskip 1em plus 0.5em minus 0.4em\relax IEEE, 2018,
  pp. 1--5.

\bibitem{kuran2011modulation}
M.~S. Kuran, H.~B. Yilmaz, T.~Tugcu, and I.~F. Akyildiz, ``Modulation
  techniques for communication via diffusion in nanonetworks,'' in
  \emph{Communications (ICC), 2011 IEEE International Conference on}.\hskip 1em
  plus 0.5em minus 0.4em\relax IEEE, 2011, pp. 1--5.

\bibitem{shahmohammadian2012optimum}
H.~ShahMohammadian, G.~G. Messier, and S.~Magierowski, ``Optimum receiver for
  molecule shift keying modulation in diffusion-based molecular communication
  channels,'' \emph{Nano Communication Networks}, vol.~3, no.~3, pp. 183--195,
  2012.

\bibitem{kim2013novel}
N.-R. Kim and C.-B. Chae, ``Novel modulation techniques using isomers as
  messenger molecules for nano communication networks via diffusion,''
  \emph{IEEE Journal on Selected Areas in Communications}, vol.~31, no.~12, pp.
  847--856, 2013.

\bibitem{gine2009molecular}
L.~P. Gin{\'e} and I.~F. Akyildiz, ``Molecular communication options for long
  range nanonetworks,'' \emph{Computer Networks}, vol.~53, no.~16, pp.
  2753--2766, 2009.

\bibitem{kuscu2016physical}
M.~Kuscu and O.~B. Akan, ``On the physical design of molecular communication
  receiver based on nanoscale biosensors,'' \emph{IEEE Sensors Journal},
  vol.~16, no.~8, pp. 2228--2243, 2016.

\bibitem{akan2009cognitive}
O.~B. Akan, O.~B. Karli, and O.~Ergul, ``Cognitive radio sensor networks,''
  \emph{IEEE network}, vol.~23, no.~4, 2009.

\bibitem{berezhkovskii2013effect}
A.~M. Berezhkovskii and A.~Szabo, ``Effect of ligand diffusion on occupancy
  fluctuations of cell-surface receptors,'' \emph{The Journal of Chemical
  Physics}, vol. 139, no.~12, p. 09B610\_1, 2013.

\bibitem{hasselblad1969estimation}
V.~Hasselblad, ``Estimation of finite mixtures of distributions from the
  exponential family,'' \emph{Journal of the American Statistical Association},
  vol.~64, no. 328, pp. 1459--1471, 1969.

\bibitem{jewell1982mixtures}
N.~P. Jewell, ``Mixtures of exponential distributions,'' \emph{The Annals of
  Statistics}, pp. 479--484, 1982.

\bibitem{kay1993fundamentals}
S.~M. Kay, \emph{Fundamentals of statistical signal processing: Estimation
  theory}.\hskip 1em plus 0.5em minus 0.4em\relax Englewood Cliffs NJ USA:
  Prentice-Hall, 1993.

\bibitem{daniel2013synthetic}
R.~Daniel, J.~R. Rubens, R.~Sarpeshkar, and T.~K. Lu, ``Synthetic analog
  computation in living cells,'' \emph{Nature}, vol. 497, no. 7451, p. 619,
  2013.

\bibitem{chou2017chemical}
C.~T. Chou, ``Chemical reaction networks for computing logarithm,''
  \emph{Synthetic Biology}, vol.~2, no.~1, p. ysx002, 2017.

\end{thebibliography}
\end{document}